\documentclass[journal]{IEEEtran}
\usepackage{xcolor,soul,framed} 
\colorlet{shadecolor}{yellow}
\usepackage{graphicx}
\graphicspath{{../pdf/}{../jpeg/}}
\usepackage[cmex10]{amsmath}
\usepackage{array}
\usepackage{mdwmath}
\usepackage{mdwtab}
\usepackage{eqparbox}
\usepackage{url}
\usepackage{bm}
\usepackage{enumerate}
\usepackage{amsfonts}
\usepackage{nomencl}
\usepackage{etoolbox}
\usepackage{amssymb}
\usepackage{mathrsfs}
\usepackage{amsmath}
\usepackage{amsfonts,amssymb}
\usepackage{float}
\graphicspath{{figures/}}
\usepackage{calc}
\usepackage{tabularx}
\usepackage{fancybox}
\usepackage{subfig}
\usepackage{xcolor}
\usepackage[ruled,vlined]{algorithm2e}
\usepackage{bm}
\usepackage{bbm}
\DeclareMathOperator*{\argmin}{arg\,min}

\usepackage{pdfpages}
\usepackage{svg}
\usepackage{booktabs}
\usepackage{tabu}
\usepackage{multirow}
\usepackage{makecell}
\usepackage{array}
\usepackage{hyperref}

\renewcommand{\baselinestretch}{0.98}
\begin{document}

\title{Cost-Oriented Load Forecasting}

\author{Jialun Zhang,
Yi Wang,~\IEEEmembership{Member,~IEEE,} 
    and Gabriela Hug,~\IEEEmembership{Senior Member,~IEEE}
        
}
\maketitle

\begin{abstract}
Accurate load prediction is an effective way to reduce power system operation costs. Traditionally, the mean square error (MSE) is a common-used loss function to guide the training of an accurate load forecasting model. However, the MSE loss function is unable to precisely reflect the real costs associated with forecasting errors because the cost caused by forecasting errors in the real power system is probably neither symmetric nor quadratic. To tackle this issue, this paper proposes a generalized cost-oriented load forecasting framework. Specifically, how to obtain a differentiable loss function that reflects real cost and how to integrate the loss function with regression models are studied. The economy and effectiveness of the proposed load forecasting method are verified by the case studies of an optimal dispatch problem that is built on the IEEE 30-bus system and the open load dataset from the Global Energy Forecasting Competition 2012 (GEFCom2012).\end{abstract}
\IEEEpeerreviewmaketitle

\section{Introduction} 
\IEEEPARstart{A}{ccurate} load prediction is an effective way to reduce the power system operation costs. Both over- and under-forecasts may result in extra operational costs. For the over-forecasting case, the extra costs can be attributed to the start-up of unnecessary units, the purchase of surplus power, and selling surplus power at an unfavorable balance price \cite{Brandenberg2017TheProblem}. For the under-forecasting case, the additional costs may be due to the sub-optimal dispatch, purchase of expensive balancing power, and load shedding \cite{Weron2014ElectricityFuture}. The economic benefit of reducing the load forecasting errors in a generic model was investigated in \cite{Ranaweera1997EconomicForecasting}. It concludes that a $5\%$ forecasting error can be set as an economically acceptable allowance for the forecasting model because a further reduction of forecasting errors does not lead to a noticeable additional economic improvement. 
The economic impact of forecasting errors was quantified using three sources in \cite{Ortega-Vazquez2006EconomicInterruptions}, namely start-up costs, dispatch costs, and outage costs. In \cite{Delarue2008AdaptiveForecasting}, the economic value associated with limited and inaccurate load forecasts in a specific time frame was determined in a unit commitment problem. 

A significant amount of work has been done in the area of load forecasting, which includes both deterministic forecasting \cite{KUSTER2017257,Hong0000,Jingrui7021896} and probabilistic forecasting \cite{HONG2016914,Hong6595138,Jingrui7163624}. The forecasting objectives range from predicting the consumption of an individual consumer \cite{7885096} to the total aggregated load of the whole system \cite{8372953}. 
Traditionally, the regression models for deterministic load forecasting are trained using the metric of mean square error (MSE) loss function, with the implication that forecasting errors, identical in magnitude, cause the same and quadratic costs. Apparently, this assumption is not accurate, especially for cases where asymmetric costs are observed for forecasting errors in the same magnitude but with opposite sign \cite{Pinson2007TradingPower,Kebriaei2011Short-termFunction,Wang2017ImprovingOperations,Fatemi1777,Croo1888}. 
Thus, the forecasting errors are not able to precisely quantify the economic value among different forecasting models. 

With the discrepancy observed between the economic value and the forecasting accuracy metrics, several works studied the possibility of incorporating cost-oriented loss functions into the regression models.
An asymmetric monetary loss function was proposed in \cite{Kebriaei2011Short-termFunction} to guide the day-ahead load forecasting. On this basis, a genetic algorithm (GA) was applied to update the parameters of the radial basis function (RBF) network with a discontinuous and non-symmetric loss function.
Similarly, an asymmetric error penalty function was designed in \cite{Wang2017ImprovingOperations} based on the simulation results from the day-ahead unit commitment (DAUC) problem. The derived non-differentiable penalty function was optimized through GA for the combined backpropagation and RBF neural networks. Nevertheless, the error measurement between the forecasted loads and the actual loads over the 24 hours was averaged, resulting in an inexact measurement of the loss function. 
In addition to load forecasting,  a general cost-oriented wind power forecasting model was formulated in \cite{Li2018TowardGeneration} by integrating a predefined loss function with a boosted regression tree model. However, the loss function was designed based on a simple retail market which likely does not scale to complex scenarios. 
Another asymmetric loss function for solar power forecasting was studied in \cite{Khabibrakhmanov2016OnErrors}, where the loss function was formulated by adding a linear term and a cubic term as perturbations to the MSE loss function. The coefficients of the two additional terms control the degree of asymmetry of the loss function. Nonetheless, the asymmetric loss function was still defined from a statistical perspective and may not fully reflect the real cost of the system. 

The current research on cost-oriented load forecasting is still limited, and the two following main issues are not properly addressed. First, the asymmetric loss function is generally defined for only one specific application, and there is no general framework or approach to obtain the cost-oriented loss function. Second, differentiability is not incorporated as an important characteristic for the cost-oriented loss function, which limits the usage of the respective loss functions with traditional regression models. 
To address these two issues, this paper presents a generalized cost-oriented load forecasting framework that can be divided into the three stages of loss function generation, approximation, and integration into regression models. The losses caused by forecasting errors are first generated by a large number of simulations. On this basis, a differentiable piecewise loss function is mathematically approximated and derived. Finally, the differentiable piecewise loss function can be integrated with all regression models that can be trained by gradient-based optimization methods. 

\begin{figure*}[t]
\centering
\includegraphics[width=0.68\linewidth]{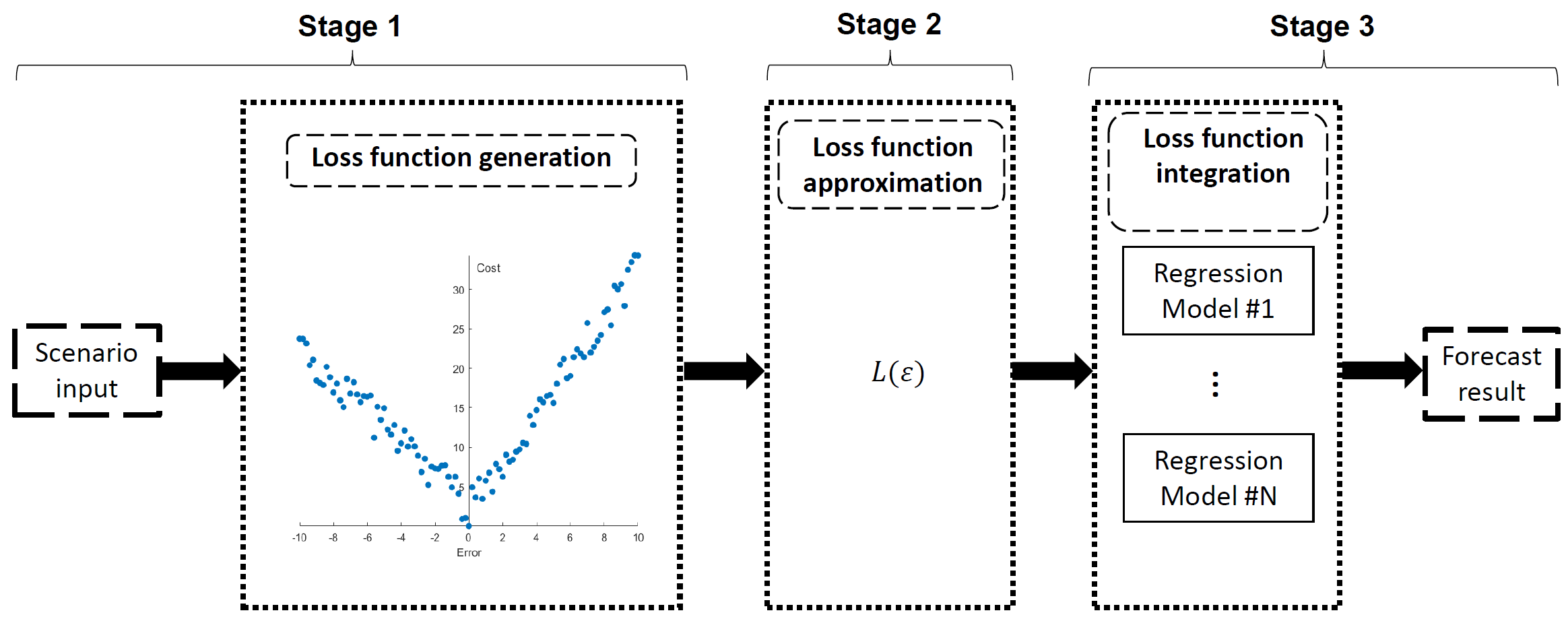}
\caption{Solution framework for cost-oriented load forecasting.}
\label{fig:framework}
\end{figure*}

Hence, this paper makes the following main contributions:
\begin{enumerate}
    \item A generalized framework to quantify losses with respect to different forecasting errors is proposed. Specifically, an example model is formulated to generate the loss data associated with forecasting errors by comparing the day-ahead economic dispatch (DAED) problem and the intraday power balance (IPB) process. 
    \item A fully differentiable cost-oriented loss function is derived by applying an optimal piecewise linear approximation method and the Huber norm embedding technique on the generated loss data.
    \item The cost-oriented loss function is integrated with multiple linear regression (MLR) and artificial neural network (ANN) models that are trained under gradient-based optimization methods. 
\end{enumerate}

The rest of the paper is organized as follows. Section~\ref{Statement} identifies the critical issues in cost-oriented load forecasting and provides a general solution framework. In Section~\ref{Forloss}, the process to generate losses with respect to different forecasting errors and to derive the differentiable cost-oriented loss function from the discrete loss data is illustrated. Section~\ref{Forecasting} shows how to integrate the derived loss function with MLR and ANN models to formulate forecasting models. Section~\ref{Casestudies} presents case studies to verify the effectiveness of the proposed method based on a DAED-IPB problem. Section~\ref{conclusion} draws conclusions.

\section{Problem Statement and Framework}\label{Statement}
Traditional load forecasting models try to minimize the MSE or other statistical indices. In contrast, the cost-oriented load forecasting model discussed in this paper tries to minimize the real economic costs caused by forecasting errors in the decision-making process. In general, the loss function and regression model are two parts that form the basis for a load forecasting model. Thus, to develop a cost-oriented load forecasting model, we need to address two main challenges. The first is to define a loss function that can reflect the real cost of power system operation. This is necessary since the economic costs caused by forecasting errors may vary in different situations and time periods. For example, a 3\% load forecasting error at a certain time on different days may yield different costs due to the varying operating conditions of the power system. The second challenge is to integrate the cost-oriented loss function with a regression model. The mature and off-the-shelf regression models use quadratic loss functions (e.g., MSE). Hence, it is necessary to study how to replace the quadratic loss function with the cost-oriented loss function and meanwhile guarantee that the model can be easily trained.

Thus, the proposed framework for the cost-oriented load forecasting model is shown in Fig.~\ref{fig:framework}. The framework consists of the following three stages:

\begin{enumerate}
    \item Stage 1: Loss function generation. This stage quantifies the economic costs in the energy system associated with forecasting errors using a large number of scenarios.
    \item Stage 2: Loss function approximation. In this stage, an analytical and differentiable cost-oriented loss function based on the discrete cost and error data obtained in Stage 1 is formulated.
    \item Stage 3: Loss function integration. This stage integrates the differentiable cost-oriented loss function obtained in Stage 2 with various regression models that are trained by gradient-based optimization methods. 
\end{enumerate}

In the following Section~\ref{Forloss}, we demonstrate how to generate and approximate the loss function before then integrating it into regression models in Section~\ref{Forecasting}. 

\section{Formulation of loss function}\label{Forloss}
\subsection{Loss Function Generation}\label{Scenario}
To quantify the costs associated with forecasting errors, two costs, namely the ideal cost with accurate forecasts and the actual cost with forecasting errors, need to be computed. For the ideal cost, the system operator gathers all the precise information of the system ex-ante such that the information $I$ available at current time $i$ is the same as at the instance of time $i+k$ for which the prediction has been made, namely $I_i^*=I_{i+k}$. Given the exact information, namely using the load forecasts at future time $y_{i+k}^*$, the system operator can obtain the minimum system operation cost $C(y_{i+k}^*|I_i^*)$. However, in reality the system operator can only make a decision based on the load forecasts simulated with the current existing information $I_i$ and then adjust the system accordingly when the future time arrives. In this case, the actual cost associated with the forecasts is denoted as $C(\hat{y}_{i+k}|I_i)$. By comparing against the ideal cost, the economic loss emanated from the forecasting errors can be derived as:
\begin{align}
    C(\epsilon_{i+k})=C(\hat{y}_{i+k}|I_i)-C(y_{i+k}^*|I_i^*)
\end{align}
where the forecasting error $\epsilon_{i+k}=\hat{y}_{i+k}|I_i-y_{i+k}^*|I_i^*$. 

\begin{figure}[t]
	\centering
	\includegraphics[width=0.6\linewidth]{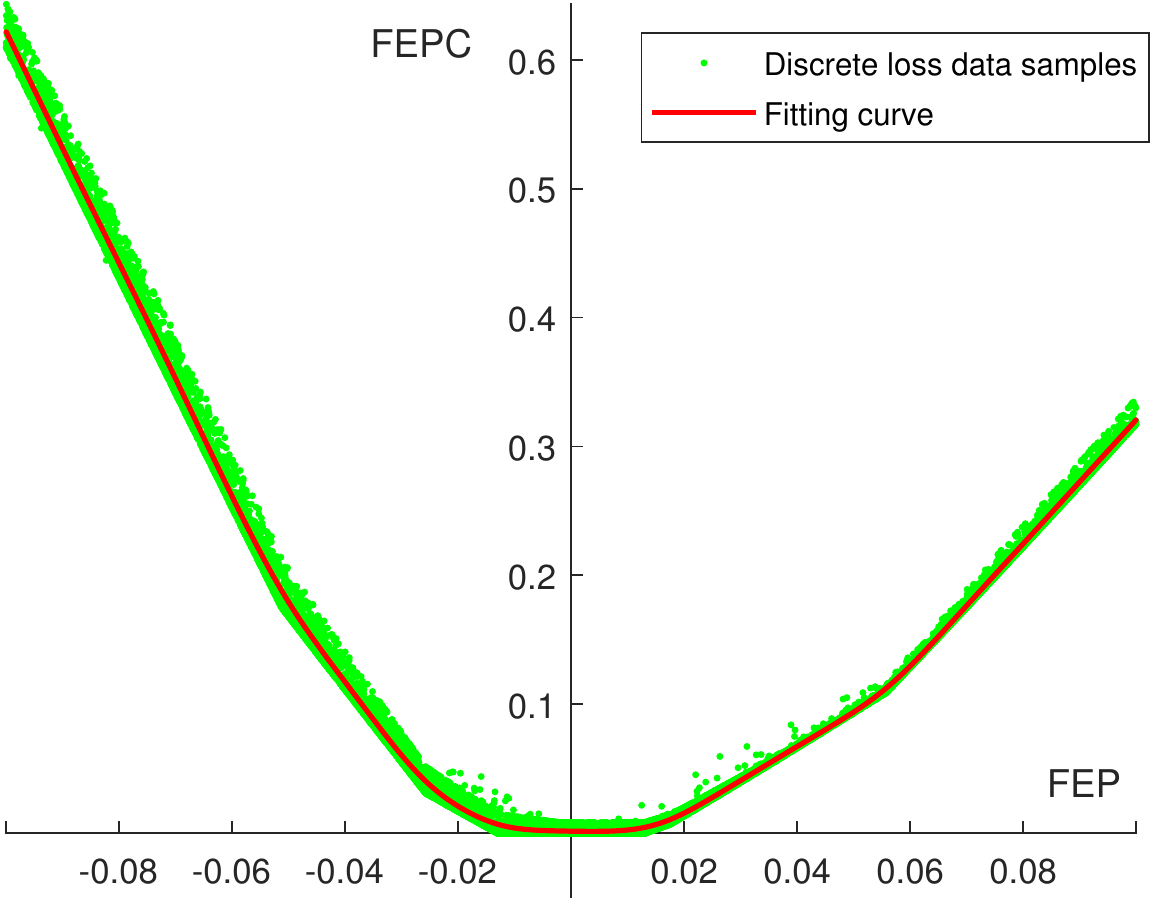}
	\caption{Smoothing spline for discrete loss data in hour 19. FEPC: Forecasting error percentage cost. FEP: Forecasting error percentage.}
	\label{fig:spline_19}
\end{figure}

Thus, by generating various scenarios of imperfect load forecasts, we can quantify the discrete economic costs associated with the simulated forecasting errors ex-ante. For example, in the case studies a massive number of load forecasts were simulated and fed into the DAED-IPB problem to minimize the total generation costs over 24 hours. On this basis, the cost $C(\epsilon_{i+k})$ for each hour is calculated. The green dots in Fig.~\ref{fig:spline_19} show the loss data simulated for hour 19, where the evaluation metrics for both costs (i.e., FEPC) and forecasting errors (i.e., FEP) are measured as normalized values from $C(\epsilon_{i+k})$ and $\epsilon_{i+k}$, respectively. 

\subsection{Loss Function Approximation}
\label{Lossfunction}
The generated loss data correspond to discrete samples. However, the loss function has to be a unique mapping between errors to losses as well as analytical over the entire spectrum such that it can be integrated into an optimization problem and solved by a gradient descent algorithm. To generate a unique loss function, the smoothing spline is first applied to the original discrete loss data. However, as illustrated in \cite{Perperoglou2019AR}, the smoothing spline consists of $N$ natural cubic splines where $N$ denotes the number of unique discrete data points. In other words, when the number of unique discrete loss data samples (i.e., the green dots shown in Fig.~\ref{fig:spline_19}) is large, the smoothing spline creates a myriad of breakpoints, making it difficult to formulate an analytical form of the loss function that then can be integrated with the regression models. A piecewise linear approximation, on the other hand, is simple, and the number of breakpoints depends on the precision requirement of the approximation. Hence, after the smoothing spline is applied on the discrete loss data, a piecewise linear approximation of the smoothing spline is introduced to form the analytical form of the loss function. In order to carry out piecewise linear approximation, in the following subsections, a partition scheme is first explained, and the fully differentiable cost-oriented loss function is developed on this basis.  

\subsubsection{Piecewise Linear Approximation}
The piecewise linear approximation function that is applied on the smoothing spline can be defined as:
\begin{equation}
L(\epsilon) =  \begin{cases} 
      a_1\epsilon+b_1, &  \epsilon_{\min} < \epsilon < \epsilon_{1} \\
      ... \\
      a_k\epsilon+b_k, & \epsilon_{k-1} < \epsilon < \epsilon_{k} \\
      ... \\
      a_K\epsilon+b_K, & \epsilon_{K-1} < \epsilon < \epsilon_{\max}
  \end{cases}\label{eq:point_8}
\end{equation}
with a limited set of breakpoints denoted as $\Phi = \left[(\epsilon_{1},L_{1}),...,(\epsilon_{k},L_{k}),...,(\epsilon_{K-1},L_{K-1})\right]$. Optimally determining the set of the breakpoints is the key for piecewise linear approximation. To fix the set of breakpoints, here we concisely recapitulate the process of how to determine the number of breakpoints $K-1$ and the position of breakpoints $\epsilon_k$ according to \cite{Berjon2015OptimalApplications}. 

The $L_2$ norm error between the derived piecewise linear function $L(\epsilon)$ and the smoothing spline function $s$ is a metric used to select the optimal set of breakpoints. To minimize the $L_2$ norm error, a partition scheme that ensures convexity of the approximated function is implemented on the smoothing spline $s$. Specifically, the $L_2$ norm approximation error based on the partition scheme is bounded by: 
\begin{align}
    \|s-L_{\Phi^{\star}}(\epsilon)\|_2 & \leq  \frac{(\int_{\epsilon_{\min}}^{\epsilon_{\max}}{s''(\epsilon)^{\frac{2}{5}}d\epsilon)^{\frac{5}{2}}}}{\sqrt{120}K^2}\label{eq:lin_24}
\end{align}
where the upper bound of the $L_2$ approximation error is associated with both the second order derivative of the smoothing spline $s''(\epsilon)$ and the number of partition segments $K$. With this relationship, the minimal number of breakpoints $K-1$ required to approximate the smoothing spline within the specified tolerance of the approximation error can be determined. 

\begin{figure}[t]
	\centering
	\includegraphics[width=0.81\linewidth]{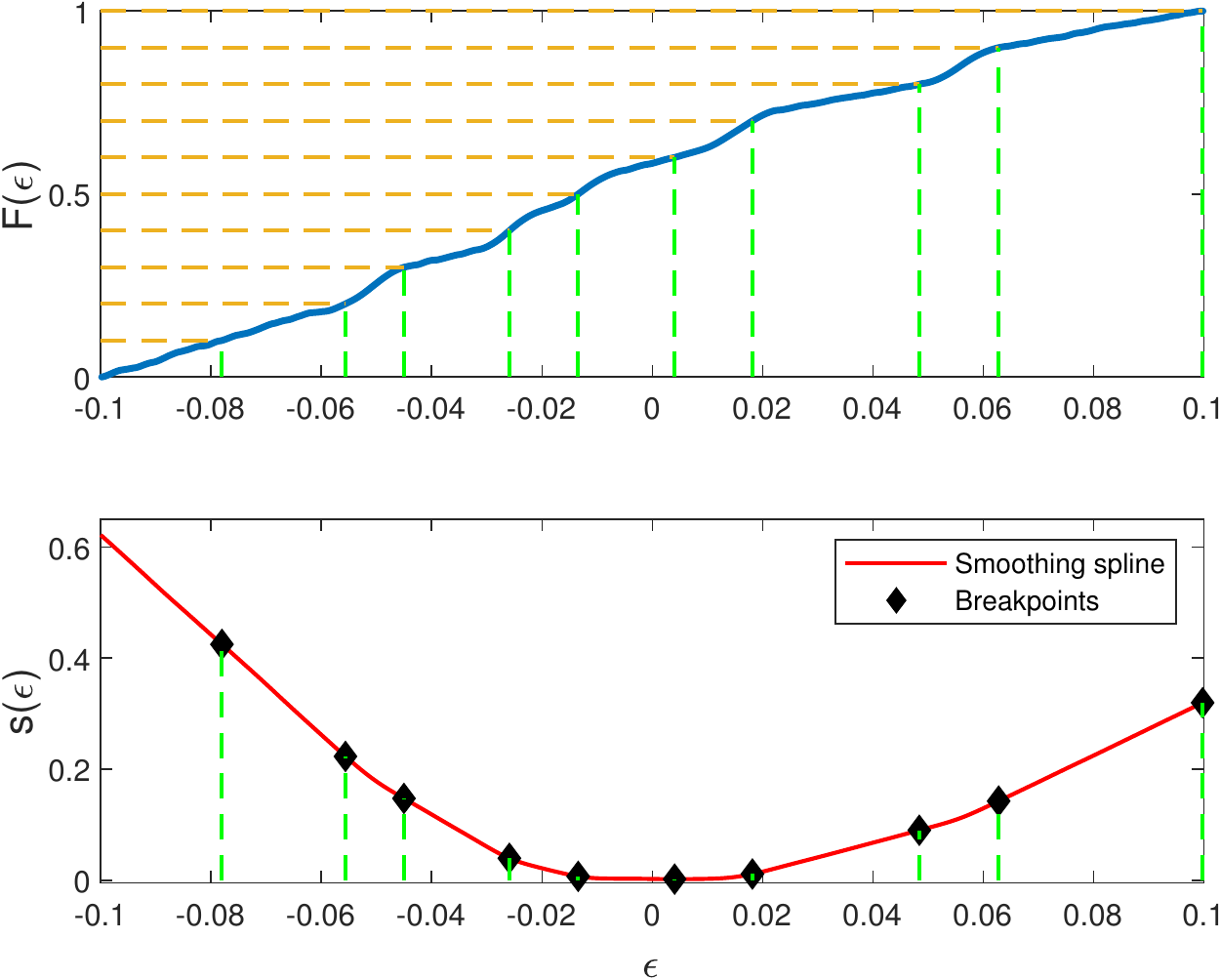}
	\caption{Partition scheme for piecewise linear approximation.}
	\label{fig:partition}
\end{figure}

Once the number of breakpoints is fixed, the next step is to determine the position of the breakpoints. 
Reference \cite{deBoor2001ASplines} proves that more breakpoints shall be placed in the subinterval where the convexity of the smoothing spline is evident. Thus, the position of the breakpoints is measured by a density metric called the cumulative breakpoint distribution function $F(\epsilon_{k})$ which is defined as:
\begin{align}
    F(\epsilon_{k}) & = \frac{\int_{\epsilon_{\min}}^{\epsilon_{k}}{|s''(x)|^{2/5}dx}}{\int_{\epsilon_{min}}^{\epsilon_{\max}}{|s''(x)|^{2/5}dx}} \label{eq:lin_18}
\end{align}
where $F(\epsilon_{k})$ is bounded by the range of $\left[0,1\right]$. Then $F(\epsilon)$ is divided evenly according to the number of the breakpoints. As Fig.~\ref{fig:partition} shows, the breakpoints $\{\epsilon_{k}\}_{k=1}^{K-1}$ are placed in accordance with the cumulative distribution function such that each subinterval can contribute equally to the total approximation error, by which the position of the corresponding breakpoints are determined.
 
\subsubsection{Differentiable Piecewise Loss Function}
After the breakpoints are determined, we can form the piecewise linear approximation of the loss data. However, since the breakpoints on both sides of each piecewise linear function create differential discontinuity, it is difficult for the loss function to be integrated with traditional regression models such as neural networks that require a continuous gradient of loss function to optimize regression coefficients. Inspired by the Huber function (a.k.a Huber norm) \cite{Huber1964RobustParameter}, the integration of transitional curves at all breakpoints are introduced to address this issue. The Huber function is defined as \cite{Huber1964RobustParameter}:
\begin{equation}
h_{\delta}(\epsilon) =  \begin{cases} 
      \frac{\epsilon^2}{2\delta} & |\epsilon| \leq \delta \\
      |\epsilon| - \frac{\delta}{2} & |\epsilon| > \delta
   \end{cases}
\end{equation}
where $\epsilon$ stands for the forecasting error and $\delta \in \mathbb{R}^{+}$ is a real value that represents the transition range from the quadratic component to the linear components.

\begin{figure}[t]
	\centering
	\includegraphics[width=0.9\linewidth]{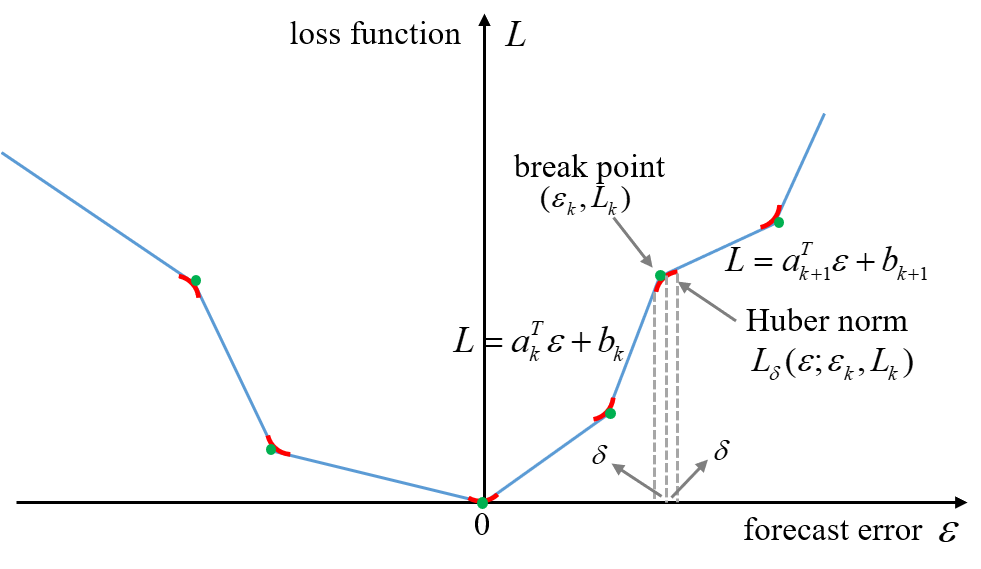}
	\caption{Continuous and differentiable piecewise linear approximation of the loss function.}
	\label{fig:huber_11}
\end{figure}
As illustrated in Fig. \ref{fig:huber_11}, a modified Huber norm is introduced in the neighborhood of each breakpoint $\epsilon_{k}$ (i.e., $[\epsilon_{k}-\delta, \epsilon_{k}+\delta]$). Now, assume that the quadratic component and two linear functions are given by:
\begin{equation}
\label{eq:hu}
\begin{aligned}
L_{H}(\epsilon) &= A\epsilon^2 + B\epsilon + C \\
L_{p}(\epsilon) = a_{k+1}\epsilon+&b_{k+1},~ 
L_{n}(\epsilon) = a_{k}\epsilon+b_k 
\end{aligned}
\end{equation}
The following four conditions have to be satisfied for the function to be differentiable: both values and the first order derivative at the transition points have to be identical, i.e.,
\begin{equation}
\begin{aligned}
&L_{H}(\epsilon_{k}+\delta) = L_{p}(\epsilon_{k}+\delta),~
L_{H}'(\epsilon_{k}+\delta) = L_{p}'(\epsilon_{k}+\delta) 
\\
&L_{H}(\epsilon_{k}-\delta) = L_{n}(\epsilon_{k}-\delta),~ 
L_{H}'(\epsilon_{k}-\delta) = L_{n}'(\epsilon_{k}-\delta)
\label{eq:hu2}
\end{aligned}
\end{equation}
Consequently, we can derive the quadratic component $L_{H}$ as follows:
\begin{equation}
\begin{aligned}
L_{H} & = \frac{a_{k+1}-a_{k}}{4\delta}(\epsilon-\epsilon_{k})^2 + \frac{a_{k+1}+a_k}{2}(\epsilon-\epsilon_{k}) \\
&+\frac{\delta(a_{k+1}-a_k)}{4} + L_k
\label{eq:ori_n}
\end{aligned}
\end{equation}
Thus, the final approximated and differentiable loss function that will be integrated into regression models is formulated as
\begin{equation}\label{eq:huberloss}
L_{\delta}(\epsilon) =  \begin{cases} 
      a_{k}(\epsilon-\epsilon_{k})+L_{k} & \epsilon_{k} - \epsilon \geq \delta \\
      \begin{split}
      &\frac{a_{k+1}-a_{k}}{4\delta}(\epsilon-\epsilon_{k})^2+ \\ &\frac{a_{k+1}+a_{k}}{2}(\epsilon-\epsilon_{k})+\\
      &\frac{a_{k+1}-a_{k}}{4}\delta + L_{k} \end{split}& |\epsilon-\epsilon_{k}| < \delta\\
      a_{k+1}(\epsilon-\epsilon_{k})+L_{k} & \epsilon - \epsilon_{k}  \geq \delta
   \end{cases}
\end{equation}

\section{Loss Function Integration}\label{Forecasting}
A load forecasting model $f(X)$ tries to establish the mapping between the relevant features $X_{n\times d}$ and the predicted loads $Y_{n\times1}$, where $d$ and $n$ denote the number of features and training samples, respectively. 
For typical deterministic forecasting models, the inherent loss function for each sample $\{x_i,y_i\}$ is the quadratic loss function $(f(x_i)-y_i)^2$. Considering the incompatibility of cost-oriented loss functions to most off-the-shelf training packages, the quadratic loss function cannot be directly replaced by the cost-oriented loss function $L(\epsilon_i)$ to train the forecasting models. Since MLR and ANN are two commonly used regression models for load forecasting \cite{Wang2016ElectricApproach}, this section studies how to train these two models with the cost-oriented loss function. In this paper, the forecasting error percentage (FEP) $\epsilon_i=\frac{f(x_i)-y_i}{y_i}$ is used as the input to the loss function $L(\epsilon_i)$ instead of the absolute error.
\subsection{MLR with Cost-oriented Loss Function}
For MLR model $f(X)=X\boldsymbol{w}$, 
given the defined piecewise loss function $L$, the optimization problem for determining the parameters $\boldsymbol{w}$ can be formulated as:
\begin{align}
  \boldsymbol{w^{*}} &= \argmin_{\boldsymbol{w}}\sum_{i=1}^n{L\left(\frac{x_i\boldsymbol{w}-y_i}{y_i}\right)}\label{eq:reg_3}  
\end{align}

It can be solved by the gradient descent method which is a process of iteratively updating the regression parameters $\boldsymbol{w}$ by following the gradient of the loss function until the gradients finally converge to zero. If the loss function is convex, the point where the gradient is equal to zero is equivalent to the global minimum solution. The update rule is:
\begin{align}
    \boldsymbol{w^{t+1}} = \boldsymbol{w^t}&-\eta^t\nabla_{\boldsymbol{w}}\sum_{i=1}^n{L\left(\frac{x_i\boldsymbol{w}-y_i}{y_i}\right)}
    \label{eq:reg_4}  
\end{align}
where 
$\eta^t$ is the learning rate at the $t$-th iteration. Initially, the learning rate $\eta^t$ is set to its default value (e.g., 0.5). To facilitate the convergence of the algorithm, the learning rate is time-variant during the training process, i.e., $\eta^{t+1}={\eta^{t}}/{(1+\gamma{t})}$, where $\gamma$ is the decay rate \cite{Bottou2010Large-ScaleDescent}. 

Using the chain rule, \eqref{eq:reg_4} is further derived as:
\begin{small}
\begin{align}
    \boldsymbol{w^{t+1}}= \boldsymbol{w^t}-\eta^t\sum_{i=1}^n{\frac{\partial L}{\partial \epsilon_i}\frac{\partial \epsilon_i}{\partial \boldsymbol{w}}}=
    \boldsymbol{w^t}-\eta^t\sum_{i=1}^n{\frac{\partial L}{\partial \epsilon_i}\frac{x_i}{y_iI^{1\times (d+1)}}}
\end{align}
\end{small}
Since $x_i$ is of the size $1\times(d+1)$, to compute the dot division, the single scalar of $y_i$ has to be propagated by an identity matrix of the same size as $x_i$, such that $y_iI^{1\times (d+1)}$ is of the same size as $x_i$.

Both initial learning rate $\eta^0$ and the decay rate $\gamma$ are hyperparameters tuned by cross-validation. Algorithm 1 summarizes the MLR model with a cost-oriented loss function.

\begin{algorithm}[t]
\SetAlgoLined
\KwIn{Loss function $L$, training data $D=(X, Y)$ where $X\in \mathbb{R}^{n\times{m}}$, initial weights $w_0 \in \mathbb{R}^{d}$, default initial learning rate $\eta^{0}=0.5$, decay rate $\gamma=0.1$, maximum iterations $t^{\max}$.}
 \While{$t\leq t^{\max}$}{
Gradient of the cost-oriented loss function: $g_t = \nabla_{\boldsymbol{w}}{\sum_{i=1}^n{L_i(\epsilon(X_i\boldsymbol{w},y_i))}}=\sum_{i=1}^n{\frac{\partial L_i}{\partial {\epsilon_i}}\frac{x_i}{y_iI^{1\times(d+1)}}}$\;
  Weight update:  $\boldsymbol{w}_{t+1} = \boldsymbol{w}_{t}-\eta^t{g_t}$\;
  Learning rate update: $\eta^{t+1}=\frac{\eta^{t}}{1+\gamma{t}}$\;
  Update iteration tag: $t = t+1$\;
}
 \KwOut{Optimized MLR model ${Y^{*}}=X\boldsymbol{w}^*$.}
 \caption{MLR with cost-oriented loss function}
\end{algorithm}


\subsection{ANN with Cost-oriented Loss Function}

For a feed-forward ANN with $Q$ hidden layers, the input data for every layer are aggregated in the vector $\boldsymbol{v}$. Thus, the initial input data $\boldsymbol{x}$ from the input layer is $\boldsymbol{v}^{(0)} = \boldsymbol{x}$. The output of the hidden layer $l=1:Q$ is:
\begin{align}
\boldsymbol{v}^{(l)} &= \phi\left(W^{(l)}\boldsymbol{v}^{(l-1)}\right)
\end{align}
where $W^l$ is the weight vector for the input of layer $l-1$ and $\phi(\cdot)$ is the activation function.

Through propagated computation of each layer, the output of the last hidden layer is passed to the output layer. Hence, the prediction $\hat{\boldsymbol{y}}$ is obtained from the output layer
 \begin{align}
     \hat{\boldsymbol{y}} &= f(\boldsymbol{x},\boldsymbol{W}) = W^{(Q+1)}\boldsymbol{v}^{(Q)}\label{eq:5.18}
 \end{align}

The cost-oriented loss function $L(\epsilon(\hat{\boldsymbol{y}},\boldsymbol{y}))$ is used to determine the forecasting loss. Consequently, the objective function of the cost-oriented loss resembles the MLR problem~\eqref{eq:reg_3} and can be defined as:
\begin{equation}
\begin{aligned}
    \hat{\boldsymbol{W}}=\argmin_{\boldsymbol{W}}{\sum_{i=1}^n{L\left(\frac{f(x_i,\boldsymbol{W})-y_i}{y_i}\right)}}
\end{aligned}
\end{equation}
The learning problem now is to find the weight matrix such that the cost-oriented loss obtained at the output layer is minimized. Since the minimization problem is a highly non-convex optimization problem, the weights matrix $\boldsymbol{W}=(W^{(1)},...,W^{(Q+1)})$ is generally optimized by gradient descent methods. However, unlike the MLR technique, whose prediction expression is explicit, the output of ANN is a nonlinear relationship of parameters in all hidden layers. Thus, the backpropagation algorithm is applied to compute the gradient with respect to the weight matrix. 

For the weight of the output layer, the chain rule can be applied. The partial derivative of the loss function with respect to prediction $\hat{y}$ is computed, such that at layer $Q$ we have
\begin{align}
    \nabla_{W^{(Q+1)}}L &= \frac{\partial{L}}{\partial{\boldsymbol{\epsilon}}}\frac{\partial{\boldsymbol{\epsilon}}}{\partial{\boldsymbol{\hat{y}}}}\frac{\partial{\boldsymbol{\hat{y}}}}{\partial{W^{(Q+1)}}}\\
    &=\delta^{(Q+1)}{(v^{Q})^{T}}
\end{align}
where $\boldsymbol{\epsilon}=\frac{\hat{\boldsymbol{y}}-\boldsymbol{y}}{\boldsymbol{y}}$, $\delta^{(Q+1)}$ is short for $\frac{\partial{L}}{\partial{\boldsymbol{\epsilon}}}\frac{\partial{\boldsymbol{\epsilon}}}{\partial{\boldsymbol{\hat{y}}}}$.

If we define $z^{(l)} = W^{(l)}\boldsymbol{v}^{(l-1)}$, the gradient of the internal weights for each hidden layer $l=Q:-1:1$ can be computed as:
\begin{align}
    \delta^{(l)} &= \frac{\partial{\phi}}{\partial{z^{(l)}}}\odot((W^{l+1})^{T}\delta^{(l+1)})\\
     &\nabla_{W^{(l)}}L = \delta^{(l)}(v^{l-1})^{T}
\end{align}
where $\odot$ denotes the pointwise multiplication operator. Using both feed-forward and backpropagation algorithms, the gradient matrix for all weights in the neural network can be updated until convergence. However, the backpropagation for all the input observations leads to high computational efforts and reduces the computational efficiency. Therefore, a stochastic gradient descent method using mini-batch \cite{Yuan2016StochasticSizes}, i.e., a subset sampling method of the whole dataset, is implemented in the proposed ANN algorithm to alleviate the parameter estimation variance while having a low per-iteration computation cost over the whole dataset. To escape local minima in the loss function, the adaptive moment (Adam) algorithm, which has an adaptive learning rate, is implemented in the regression model \cite{Kingma2015Adam:Optimization}. Adam carefully chooses the stepsizes when it updates the weight by introducing the first and second moments. Details of the weight update using Adam can be found in Algorithm 2, which summarizes the ANN with a cost-oriented loss function model.

\begin{algorithm}[t]
\SetAlgoLined
\KwIn{Loss function $L$, training data $D=(X, Y)$ where $X\in \mathbb{R}^{n\times{m}}$, number of layers $Q$, number of neuron units in layer $l$: $M_l$, maximum epochs number $ep^{max}$, random initialization weights $W_0^l \in \mathbb{R}^{M_l}$, batch size $b$ (e.g. $b = 64$), parameters setting for Adam: $\alpha=0.001,\beta_1=0.9,\beta_2=0.999,\epsilon=10^{-8}$.}
Initialize $m_0 = 0, v_0 = 0, \boldsymbol{W}_0 = 0$\;
 \While{$ep \leq ep^{max}$}{
  Randomly partition $X$ into $T=n/b$ subsets: $\{X^1,...,X^t,...,X^{T}\}$\;
  \For{$t=1$ to $T$}{
  Feed-forward computation: $f(\boldsymbol{X}^t,W_t) = W_t^{(Q+1)}\boldsymbol{v}^{(Q)}$\;
    Compute gradient of ANN using Backpropagation: $g_t = \nabla_{\boldsymbol{W}}{\sum_{j=1}^b{L(\epsilon(f(x_j^t,W),y_i}))}$\;
  Biased first moment: $m_t = \beta_1{m_{t-1}}+(1-\beta_1)g_t$\;
  Biased second moment:  $v_t = \beta_2{v_{t-1}}+(1-\beta_2)g_t^2$\;
  Unbiased first moment:  $\hat{m_t} = \frac{m_t}{1-\beta_1^t}$\;
  Unbiased second moment:  $\hat{v_t} = \frac{v_t}{1-\beta_2^t}$\;
  Weight update:  $\boldsymbol{W}_t = \boldsymbol{W}_{t-1}-\frac{\alpha\hat{m_t}}{\sqrt{\hat{v_t}+\epsilon}}$\;
  Update the subset index: $t = t+1$\;
}
Update epoch indicator: $ep = ep +1$
 }
 \KwOut{Optimized ANN model $f(\boldsymbol{x},\boldsymbol{W}^*)$.}
 \caption{ANN with cost-oriented loss function}
\end{algorithm}

\subsection{Evaluation Metrics}
Two evaluation metrics are used to quantify the performance of the forecasting algorithm. The first is the Mean absolute percentage error (MAPE):
\begin{align}
\text{MAPE}= \frac{1}{N}\sum_{i=i}^N{|\frac{\hat{y}_{i}-y_{i}}{y_{i}}|}= \frac{1}{N}\sum_{i=i}^N{|\epsilon_i|}\times 100\%\label{eq:MAPE}
\end{align}
where $N$ is the total number of test data samples. MAPE is a common-used metric to evaluate the average magnitude of forecasting errors from the statistical perspective. 

The other metric is to evaluate the operational cost caused by forecasting errors. The forecasting error percentage cost (FEPC) for each time period is defined as the scale-independent loss given the forecasting error $\epsilon_i$, namely
\begin{align}
\text{FEPC}(\hat{y}_{i},y_{i})={\frac{C(\hat{y}_{i})-C(y_{i})}{C(y_{i})}}\times 100\%\label{eq:FEPC}
\end{align}

On this basis, the mean forecasting error percentage cost (MFEPC) is defined to measure the average of the FEPC cost on the test dataset with $N$ samples: 
\begin{align}
\text{MFEPC} = \frac{1}{N}\sum_{i=i}^N{\text{FEPC}(\hat{y}_{i},y_{i})} \label{eq:MFEPC}
\end{align}  
By computing the MFEPC at the same hour of each day over a long period (e.g., a year), we can statistically assess the economical value of the forecasting model for that hour. 

Additionally, two other metrics, namely over forecasting percentage (OFP) and under forecasting percentage (UFP), are used to examine the asymmetry distribution of the forecasting errors:
\begin{align}
    \text{OFP} &=  \frac{\sum_{i=i}^N{\max((\hat{y}_i-y_i)/|\hat{y}_i-y_i|,0)}}{N}\times100\%\\
    \text{UFP} &=  -\frac{\sum_{i=i}^N{\min((\hat{y}_i-y_i)/|\hat{y}_i-y_i|,0)}}{N}\times100\%\label{eq:OFP}
\end{align}

\section{Case Studies}\label{Casestudies}
The day-ahead economic dispatch scenario in China is used in this section to verify the effectiveness and economic advantage of the proposed framework. The MLR and ANN models with different loss functions are tested. Note that the proposed cost-oriented load forecasting framework has various potential applications and is not limited to the dispatch problem.
\subsection{Experimental Setups}
\begin{figure}[t]
        \centering
        \includegraphics[width=0.8\linewidth]{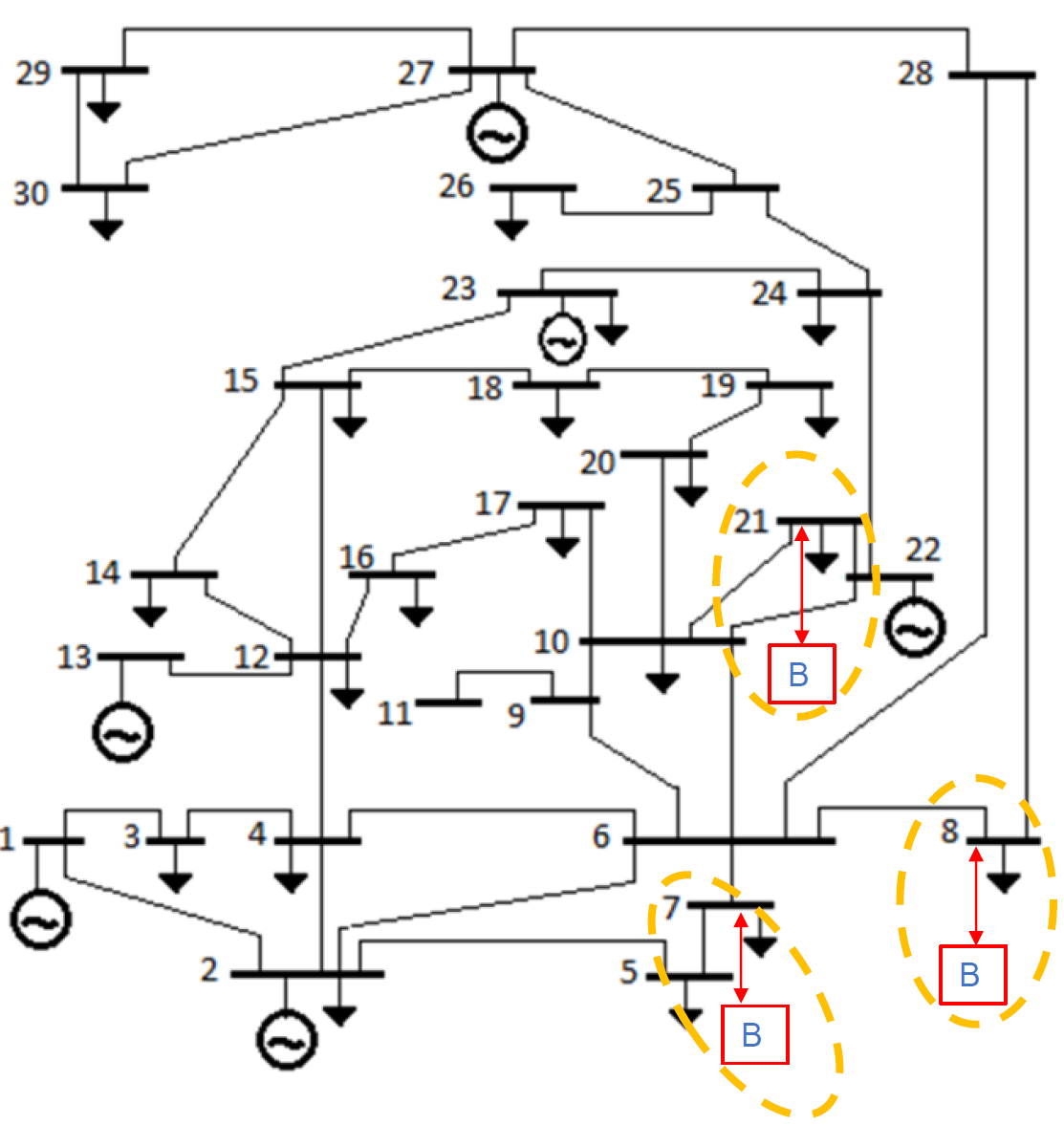}
        \caption{Topology of the 30-bus test network.}
        \label{fig:Load1}
    \end{figure}

Many provincial and municipal system operators still centrally implement the administrative generation scheduling process based on daily point forecasts  \cite{Bao8975340}. In this case, a modified IEEE 30-bus test system given in \cite{Hota2016AnalyticalSystem} is used to simulate an economic dispatch optimization problem in a regulated municipal-level grid system. Fig.~\ref{fig:Load1} shows the structure of the 30-bus test system, where a total of 6 generators and 3 BESSs are connected to the network to supply the 21 loads, whose voltage level and types are assumed to be uniform in the system. Two dispatch optimization problems, namely the DAED problem and the IPB problem, are formulated to compute the loss data. The mathematical formulations of the two problems can be found in the Appendix. Alike the actual operation process, the DAED problem schedules the power output of all online generators based on the forecasted daily load profile $\hat{y}_{i}$ in the controlled region. On this basis, the IPB problem with the actual load ${y}_{i}$ is implemented to adjust the outputs of generators, charging/discharging behavior of BESS, and load shedding to balance the intra-day load deviation. With established system parameters, the cost obtained by the IPB problem is viewed as the actual total $C(\hat{y}_{i})$. The actual total costs are then compared against the ideal total costs $C({y}_{i})$ (i.e., the DAED problem is solved with actual load ${y}_{i}$) to calculate the FEPC.
    
\begin{figure}[t]
    \centering
    \includegraphics[width=1\linewidth]{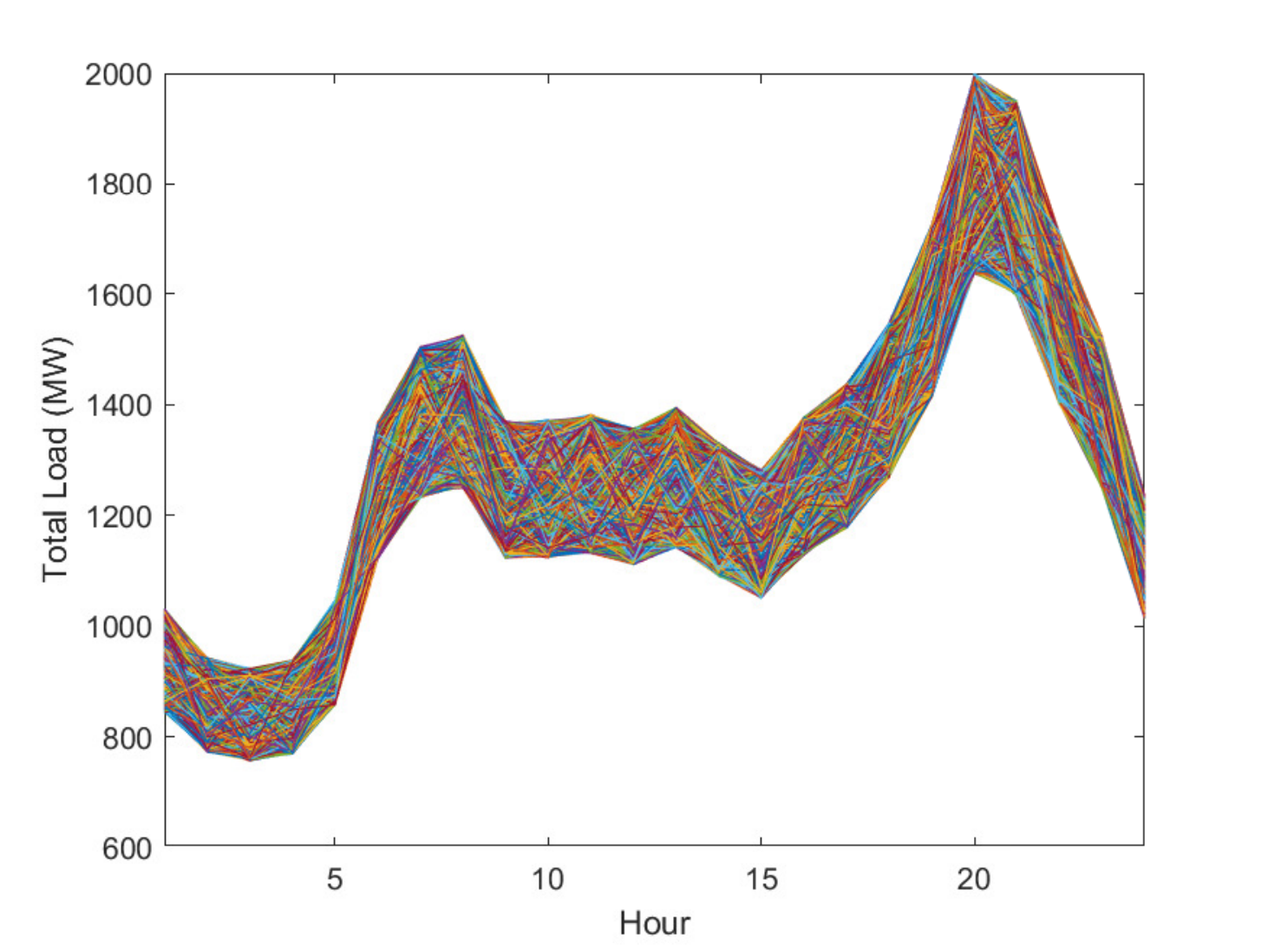}
    \caption{10,000 simulated day-ahead load forecasts.}
    \label{fig:Monte}
\end{figure}

The actual hourly load data originates from the Global Energy Forecasting Competition 2012 (GEFCom2012) dataset \cite{Hong2014Global2012} and is assumed to be distributed proportionally to each node based on the magnitude of the load provided in the modified IEEE 30-bus test case. The typical daily load data ${y}_{i}$ are derived from the original dataset to demonstrate the effectiveness of the proposed forecasting models. To simulate the forecasted loads in the day-ahead process, as shown in Fig. \ref{fig:Monte}, a total of 10,000 hourly load forecasting scenarios are generated uniformly between $\left[0.9{y}_{i},1.1{y}_{i}\right]$ in each hour $i$ using the Monte-Carlo process. Note that the goal of the simulation of forecasted loads is to generate scenarios that are widespread across the error spectrum of loss functions, which is not related to the actual distribution of the forecasting error. Since in reality only total aggregated load deviation is considered by the municipal system operators in the IPB process, the variation of uncertainty in this model is generated for the total load and simulated using the Monte Carlo process. With uniform nominal coordinate for the hourly loss functions, the same hourly loss functions can be applied to each load of the grid, thereby reducing the overall computational complexity for loss function formulation. 

Since feature selection is not the main focus of this paper, a commonly used and effective feature set that takes into account the recency effect \cite{Wang2016ElectricApproach} is applied for both MLR and ANN. For instance, the MLR model can be expressed as:
\begin{equation}
  \begin{split}
    \hat{y}_i =& q_0 + q_{1}Trend + q_{2}M_i+q_{3}W_i+q_{4}H_i+q_{5}W_iH_i\\
    &+f\left({TP}_i\right)+\sum_{ds}f\left(\tilde{TP}_{i,ds}\right)+\sum_{h}f\left({TP}_{i-h}\right)
\end{split}  
\end{equation}
where $H_i$, $W_i$, $M_i$ and $TP_i$ denote calendar hour, week, month and temperature corresponding to the hour $i$, respectively. The $f$ function comprises the coupling features between calendar variables and the temperature in hour $i$ ($TP_i$), the average temperature over the past $ds$ days ($\tilde{TP}_{i,ds}$) and the $h$ lagged hourly temperature (${TP}_{i-h}$). In our case, $ds$ is 3 and $h$ is 4. Overall, the input dataset with 1019 features is created from the data provided in GEFCom2012. 
The configuration for the ANN models integrated with different loss functions are provided in Table \ref{table:Hyperparameters}.

Four years of load data from GEFCom2012 is split into two parts for model training and testing; the first three years of data, namely from the year 2004 to 2006, are used as the training set, and the data from 2007 are used as the out-of-sample set to fairly evaluate the model performance.
\begin{table}[t]
\caption{Hyperparameters for ANN forecasting model.}
\label{table:Hyperparameters}
\begin{center}
\begin{tabular}{cc} 
\toprule
 Hyperparameter                      & Optimal value     \\\midrule

                     Number of Hidden layers             & 3                 \\
                      Number of neurons for hidden layers & (1024,2048,1024)  \\
                      Activation function                 & Relu              \\
                      Adam Learning rate                       & $10^{-3}$  \\
                     Adam parameters setting &$\beta_1=0.9,\beta_2=0.999,\epsilon=10^{-8}$ \\
                      Mini batch size & 64\\
\bottomrule
\end{tabular}
\end{center}
\end{table}

\subsection{Loss Function Generation}
\begin{figure}[t]
	\centering
	\subfloat[\label{1a}]{\includegraphics[width=0.48\linewidth]{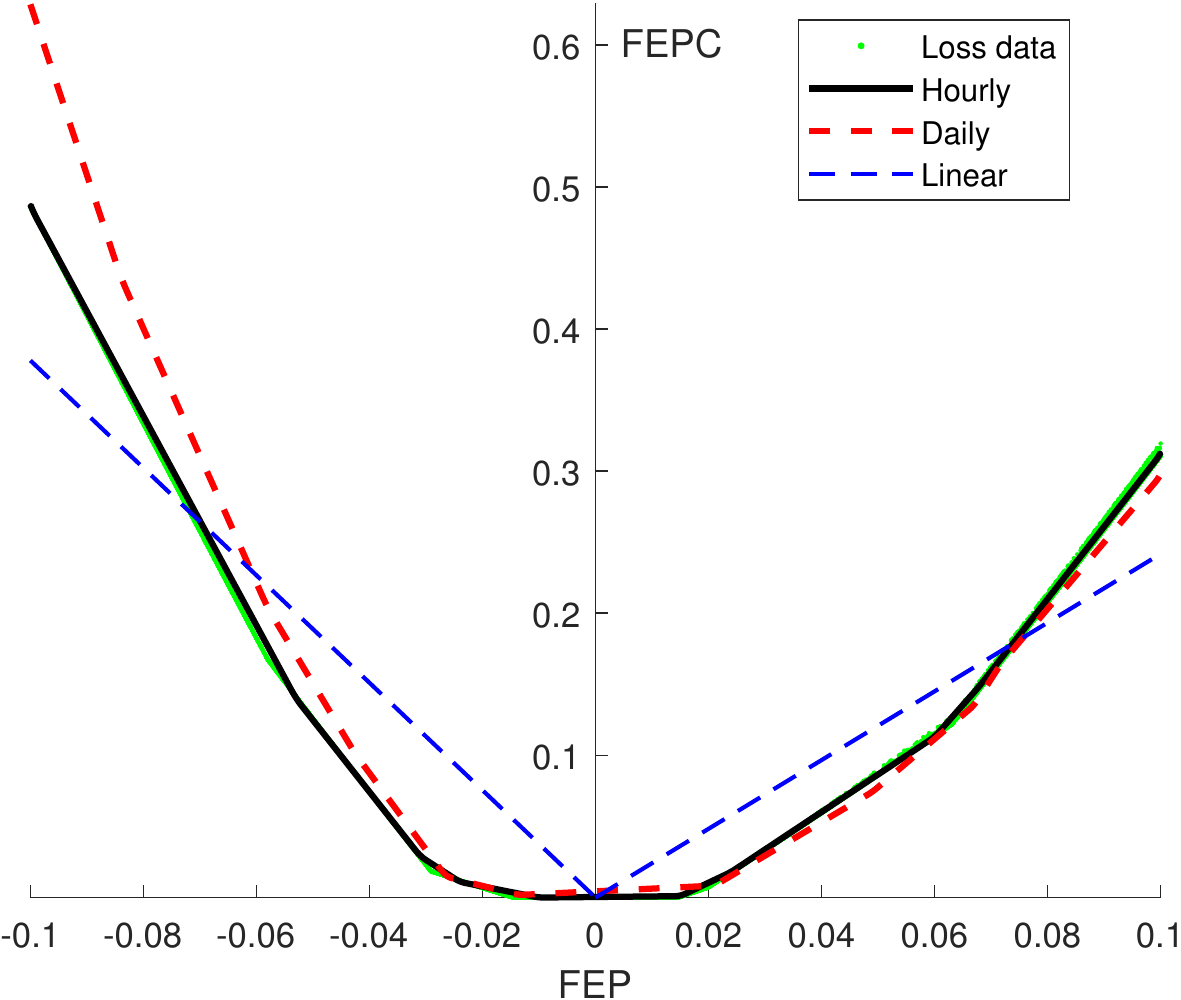}}
	\centering
	\subfloat[\label{1b}]{\includegraphics[width=0.48\linewidth]{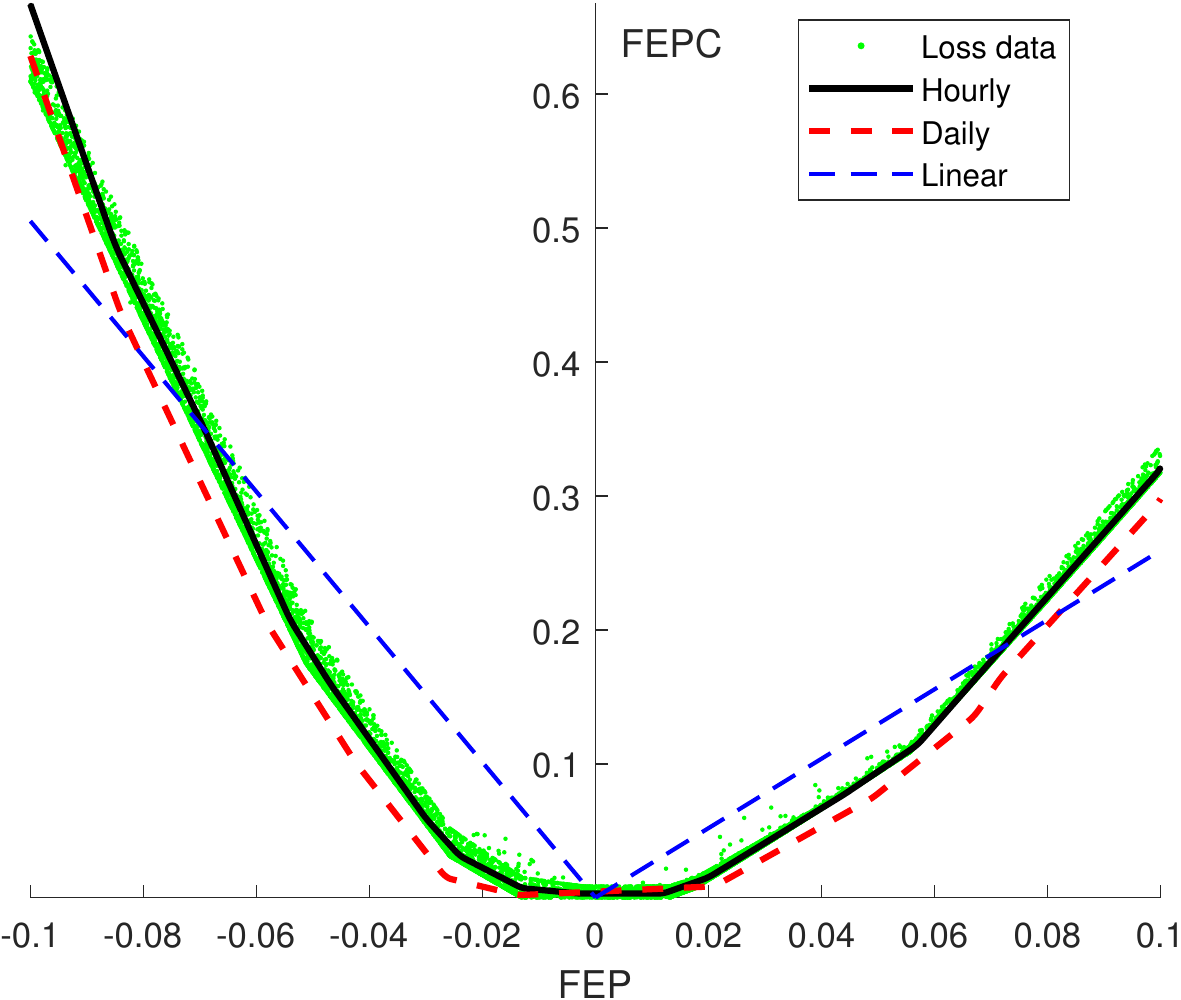}}
\caption{Cost-oriented loss functions at (a) hour 8, (b) hour 19.}
\label{fig: Loss}
\end{figure}

Fig.~\ref{fig: Loss} shows the generated loss data for hour 8 and hour 19 using DAED and IPB problems with the simulated load scenarios, where each green dot corresponds to a load scenario. The derived hourly loss data are approximated with three cost-oriented loss functions, namely hourly loss function, daily loss function, and linear loss function. The hourly loss function is derived directly from the loss data of the corresponding hour using the approximation approach proposed in Section~\ref{Forloss}. The daily loss function aggregates all the data from 24 hours and approximates them with a static loss function. The linear loss function is formulated by a linear regression model with a linear function each for positive and negative loss values.  

The approximation of the exact hourly loss data is the closest approximation to the original loss data at each hour. In this case, 24 forecasting models are trained individually for the different hours of the day. If we compare the daily loss function with the hourly loss function, we see a larger approximation error. For instance, the daily loss function is higher than the loss data when FEP is negative at hour 8, while lower than the loss data in the same range at hour 19. Nevertheless, the model integrated with the daily loss function is uniform for 24 hours which results in a lower computational burden. For the linear loss function, despite being time-dependent, it cannot approximate the real costs at each hour as precisely as the hourly loss function.

\subsection{Forecasting Results from MLR}
\begin{figure}[t]
    \centering
    \subfloat[\label{mlr_1a}]{%
        \includegraphics[width=0.45\linewidth]{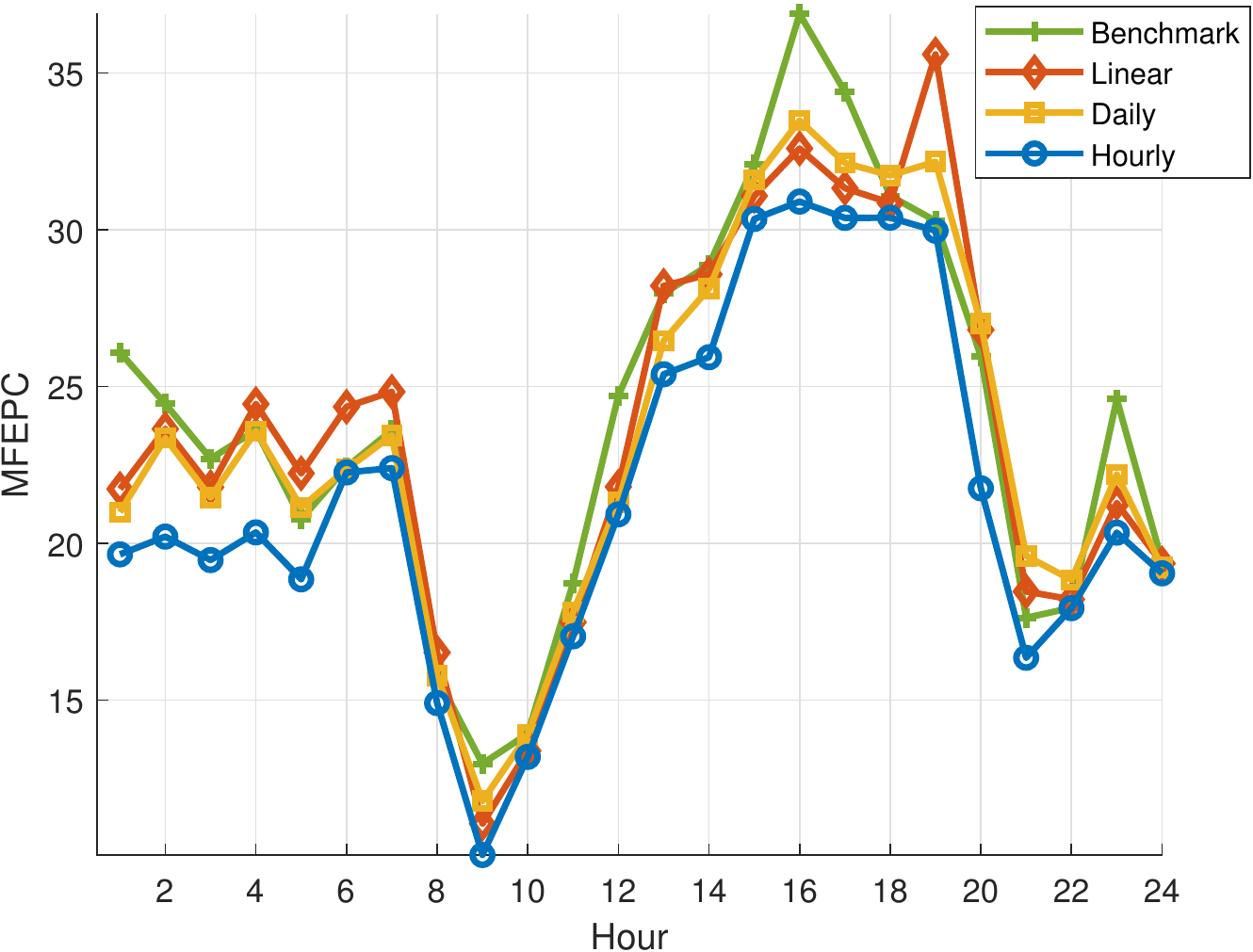}}
    \hfill
    \subfloat[\label{mlr_1b}]{%
    \includegraphics[width=0.45\linewidth]{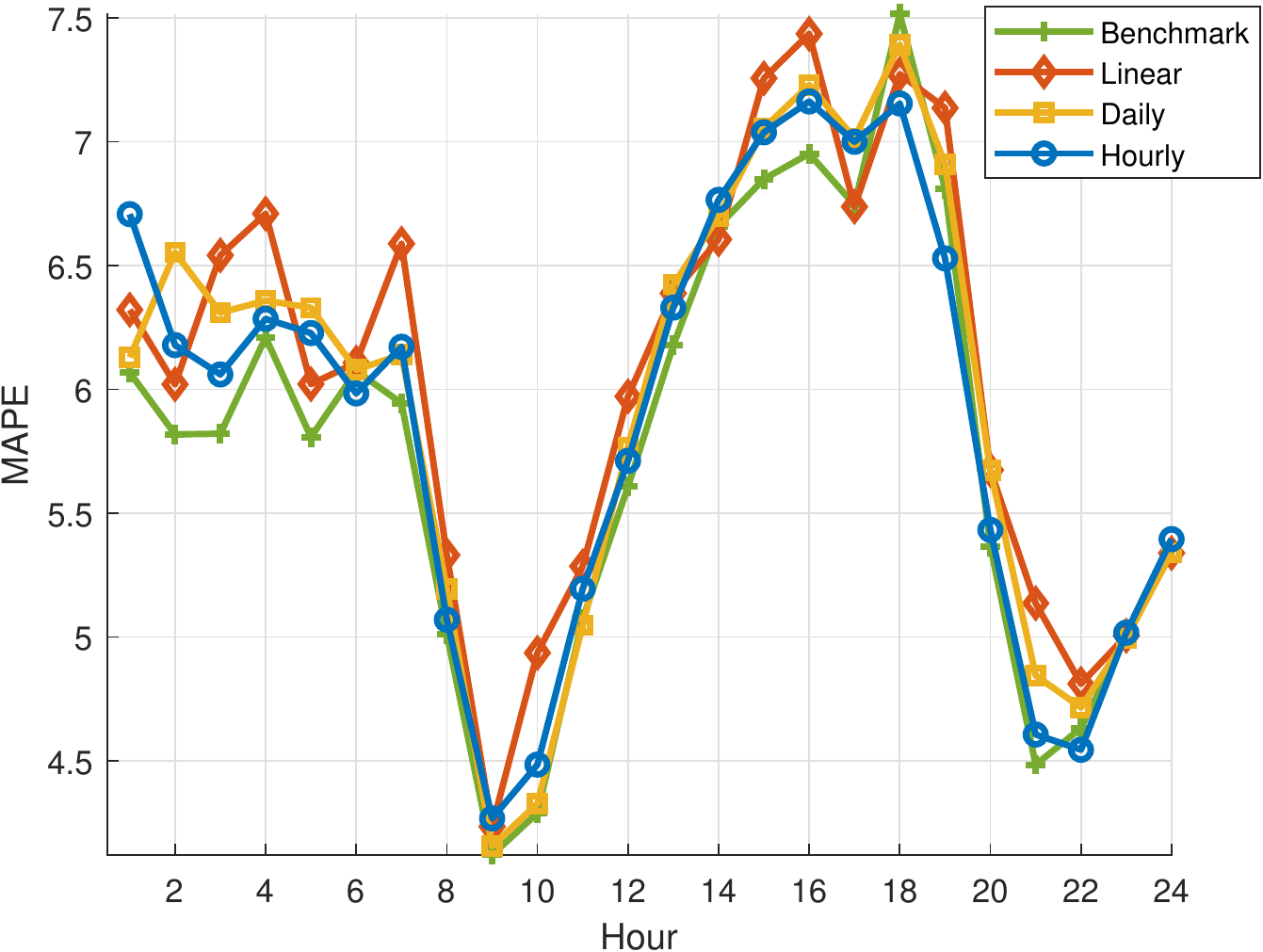}}
\\
\subfloat[\label{mlr_1c}]{%
        \includegraphics[width=0.45\linewidth]{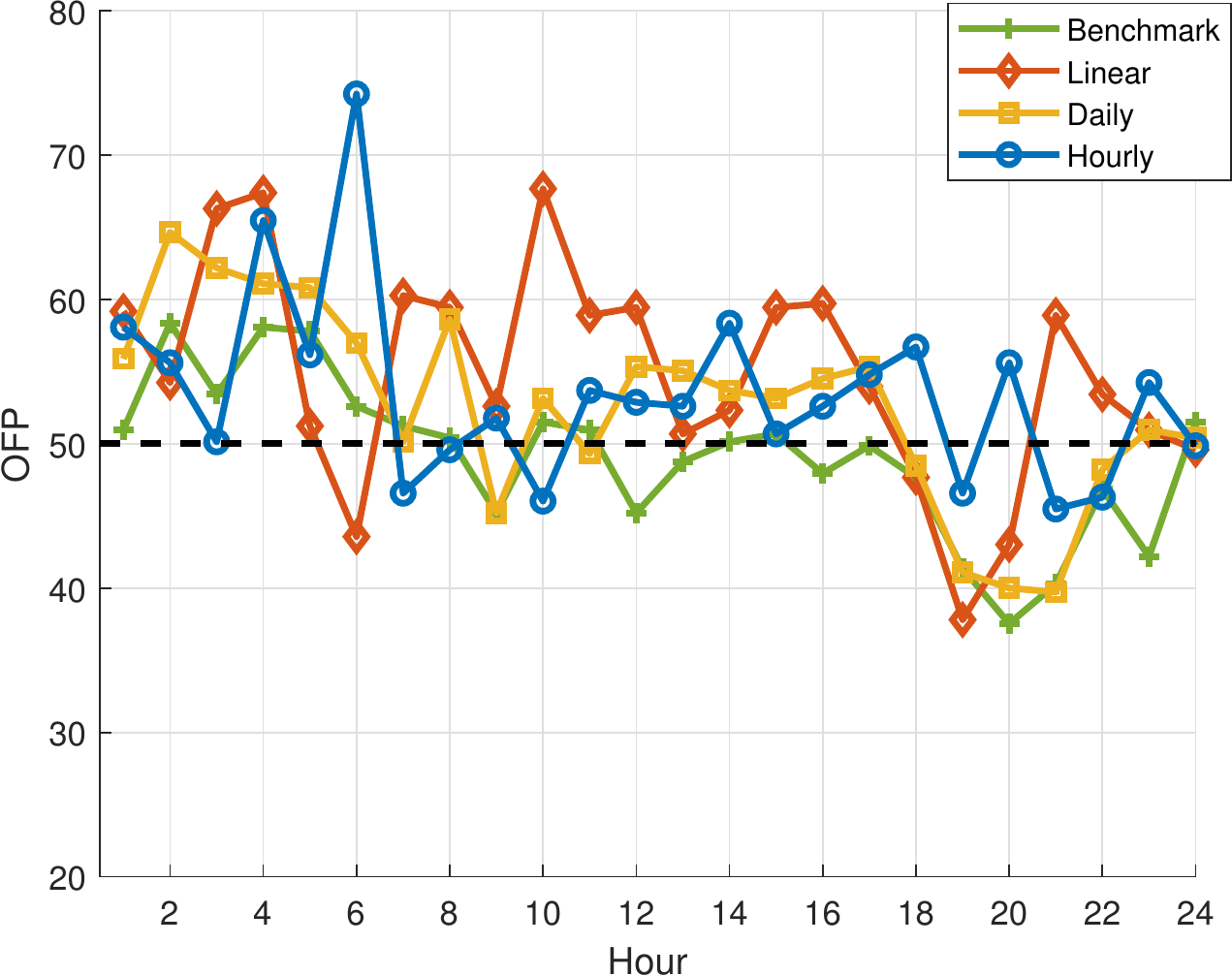}}
    \hfill
    \subfloat[\label{mlr_1d}]{%
        \includegraphics[width=0.45\linewidth]{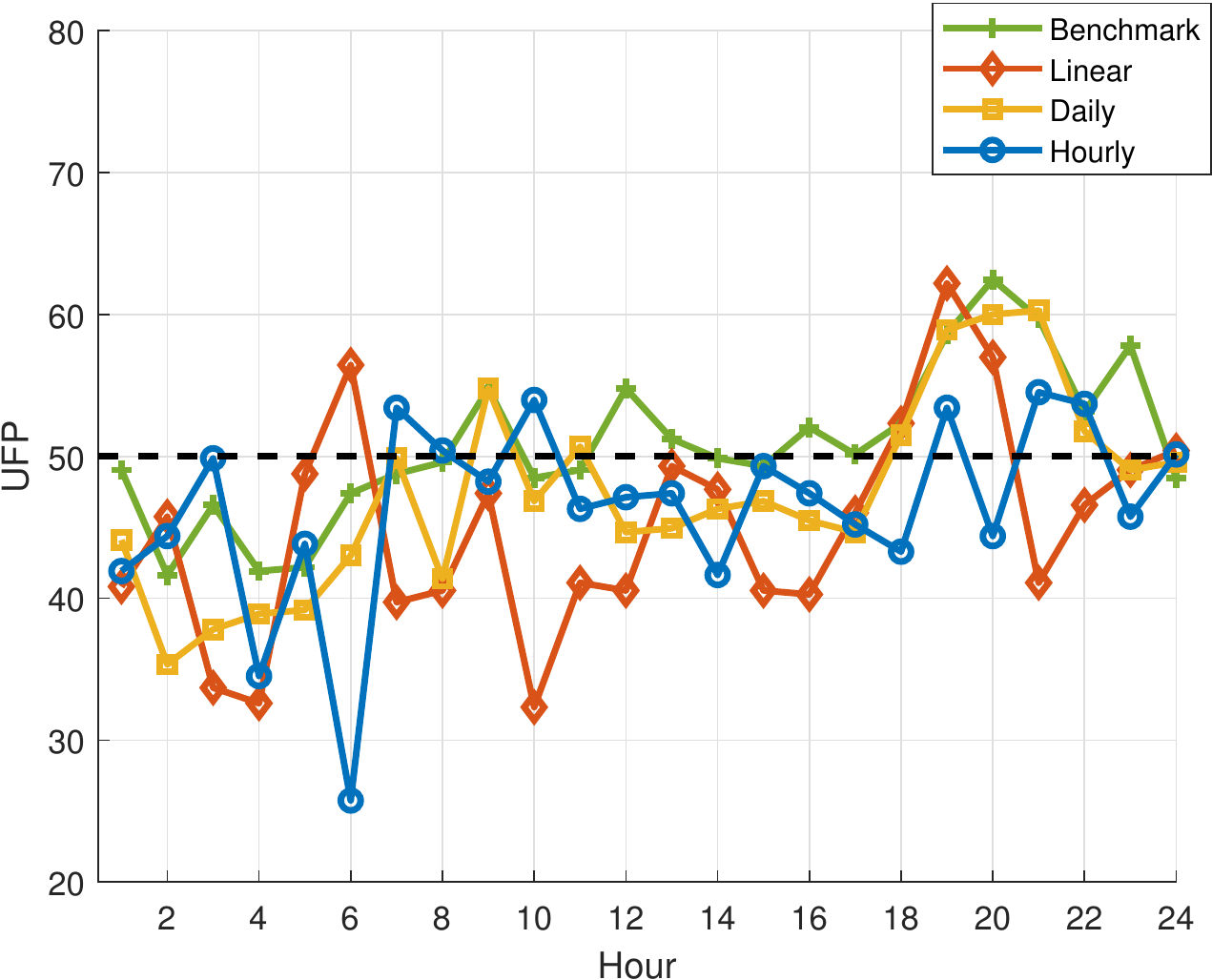}}
 \caption{\label{fig:res_mlr} Forecasting performance of four loss functions using the MLR model evaluated by metrics: (a) MFEPC, (b) MAPE, (c) OFP, (d) UFP.}
    \end{figure}

Fig. \ref{fig:res_mlr} shows the forecasting performance of MLR with the four different loss functions in terms of MFEPC, MAPE, OFP, and UFP. The forecast with the MSE loss function is used as the benchmark.
Fig.~\ref{mlr_1a} presents that the economic costs due to forecasting errors are low between the hours of 7 and 12 and high between the hours of 15 and 19 for all four loss models. A similar pattern can also be observed from the MAPE results shown in Fig.~\ref{mlr_1b}. In fact, a monotonic relationship between the forecasting errors and the economic costs can be observed for all forecasting models integrated with various loss functions. As can be expected, the economic cost at a particular hour is highly influenced by the forecasting accuracy at the same hour, which explains the correlated pattern between the MFEPC and MAPE results.

In addition, it can be seen from the load profiles in Fig.~\ref{fig:Monte} that starting from hour 7, the total load of the system is relatively constant. During this period, online generators produce a steady amount of power and are not scheduled to ramp up or down rapidly. Thus, they are relatively flexible for real-time output adjustment. In addition, the remaining energy stored in the BESSs is still abundant at the beginning of the day for balancing the load deviation. Thus, these available system resources at this period contribute to the low economic costs associated with the forecasting errors. When the total load of the system increases starting in hour 15, the system constraints are tightened, and balancing resources become limited. Thus, more costs are expected as the system is moving further away from the optimal operation point. In this case, the MFEPC associated with forecasting errors increases.

\begin{table}[t]
\renewcommand{\arraystretch}{1.6}
\caption{Daily averaged performances of four MLR models.}
\label{table:MLR_average}
\centering
\begin{minipage}{\linewidth}
\begin{center}
\begin{tabular}{lcccc}
\toprule
       & MFEPC & MAPE & OFP   & UFP    \\\hline
Benchmark & 24.04 & \textbf{5.77} & 49.19 & 50.81  \\
Linear & 23.57 & 6.04 & 54.91 & 45.09  \\
Daily  & 23.31 & 5.94 & 52.67 & 47.33  \\
Hourly & \textbf{21.59} & 5.89 & 53.50 & 46.50 \\
\bottomrule
\end{tabular}
\end{center}
\end{minipage}
\end{table}


For all loss models, the lowest MFEPC is obtained by the hourly loss model for all 24 hours. Furthermore, both daily loss and linear loss models achieve economically more favorable results over 24 hours compared to the benchmark model. Thus, the economic value of using cost-oriented forecasting models is demonstrated. On the other hand, the improvement of economic value from the daily and linear models is minor with respect to the benchmark model. This can be attributed to the simplified approximation approaches for both loss functions from the loss data, i.e., part of the information present in the hourly loss function is lost. Therefore, a trade-off between the approximation complexity of the loss function and the improvement in economic terms is observed. Overall, the MFEPC of hourly, daily, and linear models improves by 10.19\%, 3.04\%, and 1.96\% with respect to the benchmark model, respectively. 

Figs.~\ref{mlr_1c} and~\ref{mlr_1d} show that the distribution of forecasting errors for the benchmark model is almost symmetric, indicated by its metrics OFP and UFP that stay around 50\%. Nevertheless, forecasting results from the three cost-oriented loss models are biased compared to the benchmark model. This can be attributed to the asymmetric cost-oriented loss functions. In other words, the cost-oriented forecasting models impose an unbalanced economic penalty for forecasting errors with the same magnitude but with opposite signs. Specifically, the same magnitude of FEP in the negative direction causes higher FEPC than if it is a positive deviation. In this case, cost-oriented forecasting models are trained to be over-forecasting as opposed to under-forecasting. Thus, forecasting errors from three cost-oriented models have generally higher OFP and meanwhile lower UFP than the benchmark model, which can be further verified in Table~\ref{table:MLR_average}.


\subsection{Forecasting Results from ANN}
\begin{figure}[t]
    \centering
    \subfloat[\label{ANN_1a}]{%
        \includegraphics[width=0.45\linewidth]{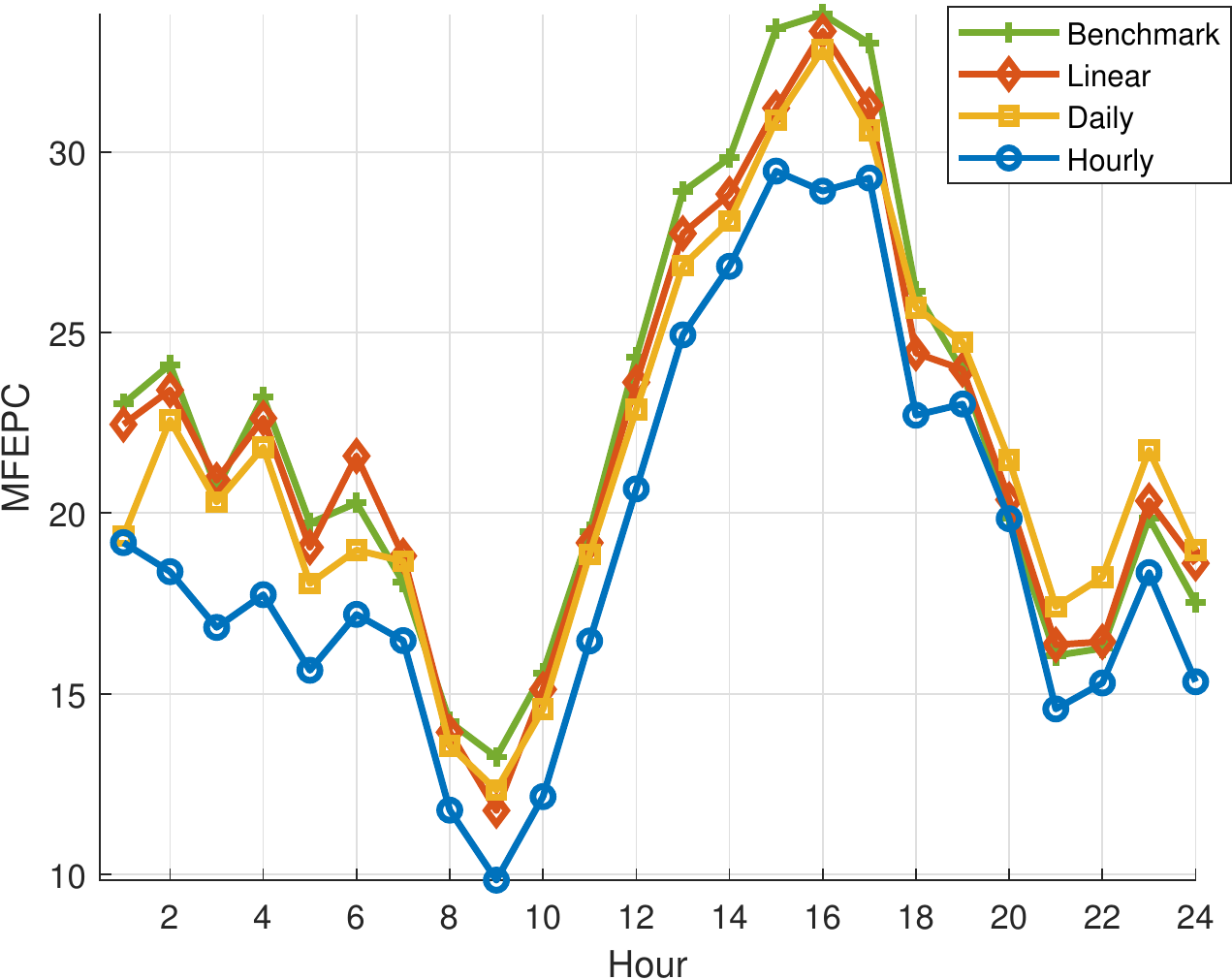}}
    \hfill
    \subfloat[\label{ANN_1b}]{%
    \includegraphics[width=0.45\linewidth]{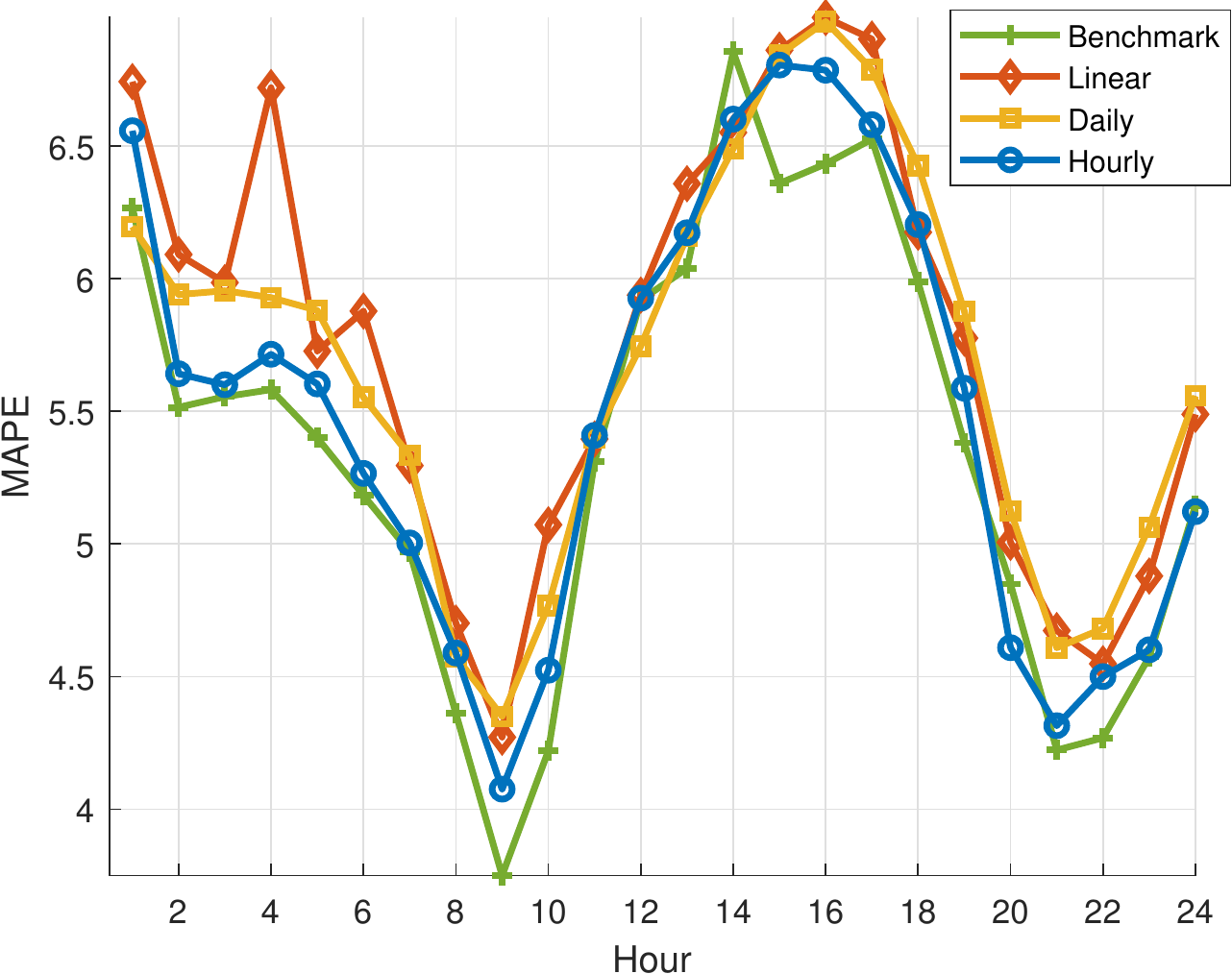}}
\\
\subfloat[\label{ANN_1c}]{%
        \includegraphics[width=0.45\linewidth]{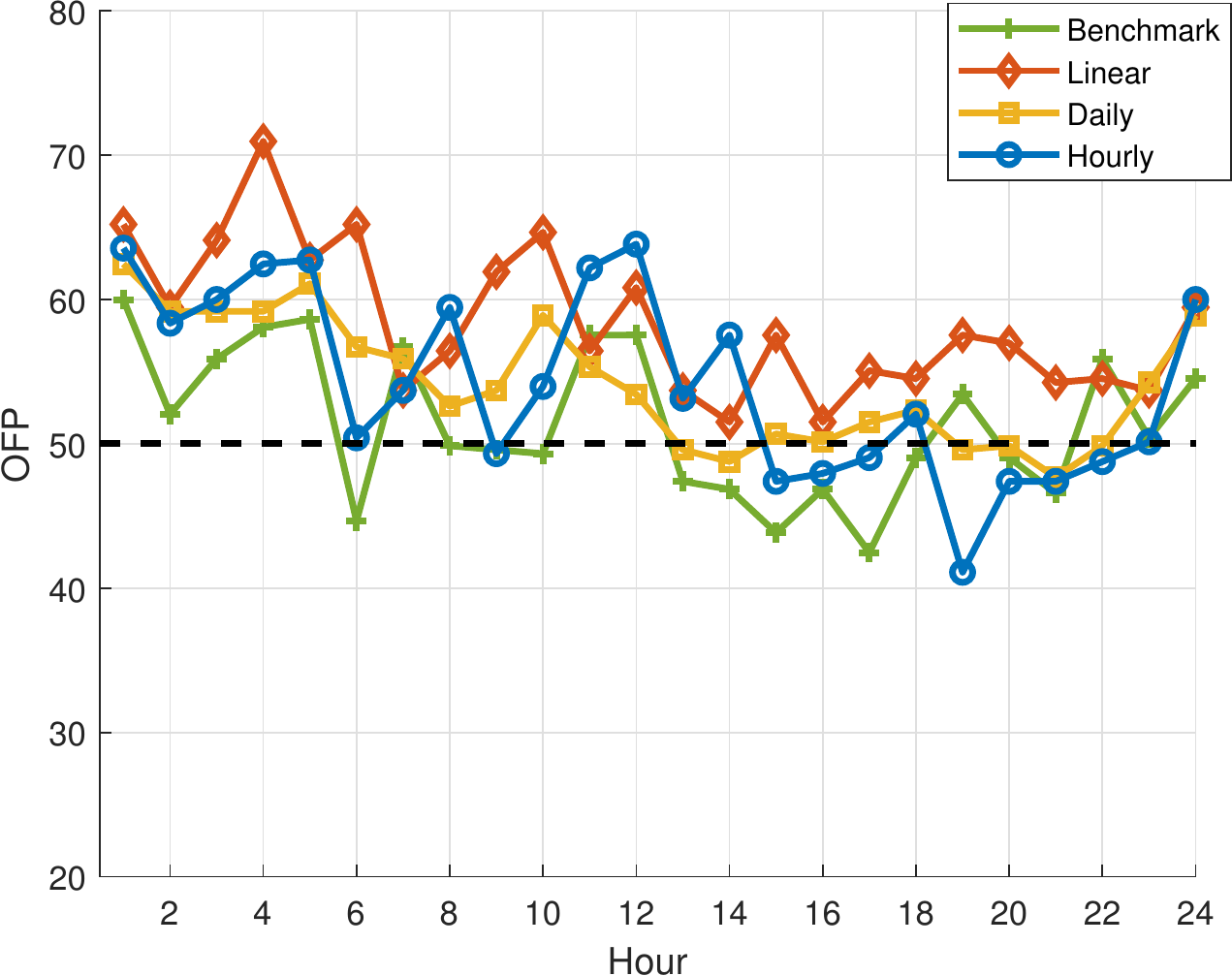}}
    \hfill
    \subfloat[\label{ANN_1d}]{%
        \includegraphics[width=0.45\linewidth]{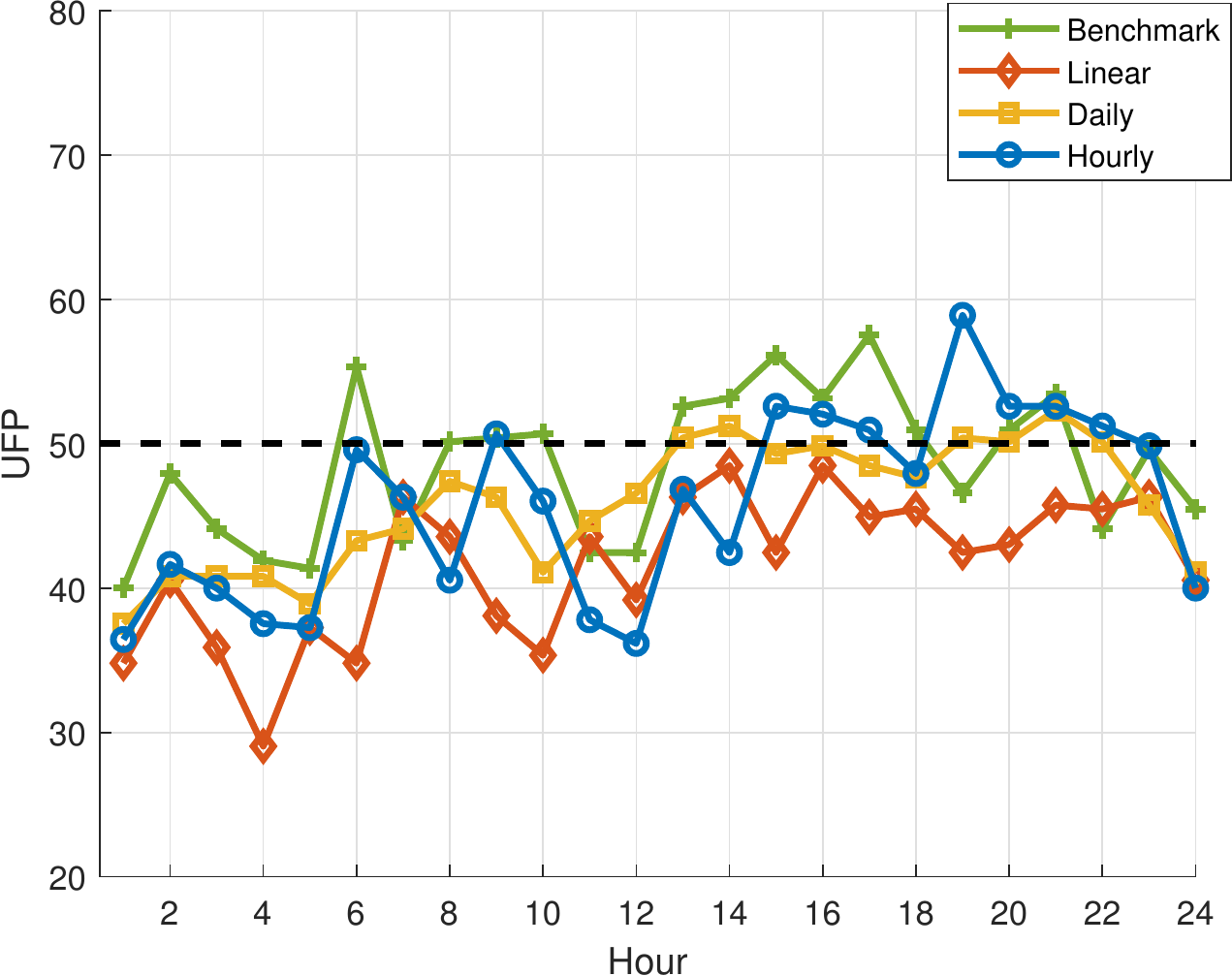}}

 \caption{\label{fig:res_ANN} Forecasting performance of four loss functions using ANN model evaluated by metrics: (a) MFEPC, (b) MAPE, (c) OFP, (d) UFP.}
    \end{figure}

Fig. \ref{fig:res_ANN} shows the forecasting performance of ANN with the four different loss functions in terms of MFEPC, MAPE, OFP, and UFP. Similarly, the hourly model using ANN obtains the lowest economic cost at all hours among the four models. Compared to MLR, the MFEPC margin between the hourly model and other loss models further increases. Table~\ref{table:ANN_average} shows that the MFEPC from hourly, daily, and linear models improves by 13.74\%, 2.83\%, and 1.71\% with respect to the benchmark model, respectively. The additional economic advantage can be largely attributed to the lower MAPE obtained by ANN models, as shown in Fig.~\ref{ANN_1b}. In other words, by improving the forecasting accuracy, ANN models obtain additional economic benefits. 
\begin{table}[t]
\renewcommand{\arraystretch}{1.6}
\caption{Daily averaged performances of four ANN models.}
\label{table:ANN_average}
\centering
\begin{minipage}{\linewidth}
\begin{center}
\begin{tabular}{lcccc}
\toprule
       & MFEPC & MAPE & OFP   & UFP    \\\hline
Benchmark & 22.27 & \textbf{5.36} & 51.51 & 48.49  \\
Linear & 21.89 & 5.75 & 58.41 & 41.59  \\
Daily  & 21.64 & 5.68 & 54.20 & 45.80  \\
Hourly & \textbf{19.21} & 5.49 & 54.25 & 45.75 \\
\bottomrule
\end{tabular}
\end{center}
\end{minipage}
\end{table}

For both MLR and ANN cases, despite having higher forecasting errors, all three cost-oriented models outperform the benchmark models economically. Among the cost-oriented loss models, hourly models achieve the highest economic gains. Due to the unbalanced loss functions, forecasting errors generated from cost-oriented loss models are more biased towards positive deviations compared to the benchmark models.  

\section{Conclusions}\label{conclusion}
This paper proposes a generalized cost-oriented load forecasting framework that is applicable to various load forecasting applications. The proposed framework enables the forecasting model to minimize the actual economic cost caused by forecasting errors. In the test cases, a heuristic model that combines DAED and IPB problems is used to generate loss data. Three differentiable cost-oriented loss functions, namely hourly loss function, daily loss function, and linear loss function, are produced from the loss data and are then integrated with MLR and ANN to form the cost-oriented forecasting models. The forecasting results from the test dataset show promising performance for the cost-oriented models. Particularly, the ANN model integrated with hourly loss function outperforms all other models in terms of the economic benefits, with up to 13.74\% improvement compared with the benchmark model trained by the traditional MSE loss functions. Meanwhile, our test case proves the feasibility of integrating customized cost-oriented loss functions with various types of load forecasting models. 



\section*{Appendix}
Here, we present the DAED and IPB combined model that is used to quantify the economic costs associated with load forecasting errors. By comparing the total cost difference between both problems, the economic value (a.k.a. loss) of forecasting errors is determined. The underlying mathematical optimization problems are formulated respectively as follows:
\subsubsection{DAED model}
\begin{align}
\min_{p_{ji},f_{\ell{i}},\delta_{ki}} \quad &\sum_{i=1}^{N}\sum_{j=1}^{J}{a_{j}p_{ji}^{2}+b_{j}p_{ji}+c_{j}} \label{eq:objective}\\
  \sum_{j=1}^{J}{M_{k}^{I}p_{ji}} &=\hat{y}_{ki} + \sum_{\ell=1}^{V}{A(\ell,k)f_{\ell{i}}} \quad : \forall{k}, \forall{i}, \label{eq:balance} \\
 & 0 \leq p_{ji} \leq \bar{p_{j}} \quad : \forall{j}, \forall{i},\label{eq:active}\\
& p_{ji} - p_{j(i-1)} \leq R_{j}^{U} \quad : \forall{j}, \forall{i \geq {2}},\label{eq:run}\\ 
& p_{ji} - p_{j}^{ini} \leq R_{j}^{U} \quad : \forall{j}, {i=1},\label{eq:ruinin}\\ 
& p_{j(i-1)} - p_{ji} \leq R_{j}^{D} \quad : \forall{j}, \forall{i \geq {2}},\label{eq:rdn}\\
& p_{j}^{ini} - p_{ji} \leq R_{j}^{D} \quad : \forall{j}, {i=1},\label{eq:rdinin}  \\  
& f_{\ell{i}} = B_{\ell}\sum_{k=1}^{K}{A(\ell,k)\delta_{ki}} \quad : \forall{\ell}, \forall{i}, \label{eq:flowing}\\
-&\bar{f}_{\ell} \leq f_{\ell{i}} \leq \bar{f}_{\ell} \quad : \forall{\ell}, \forall{i}, \label{eq:transing}    \\
-&2\pi \leq \delta_{ki} \leq 2\pi \quad : \forall{k}, \forall{i}, \label{eq:ang}\\
& \delta_{{k_{1}}i} = 0 \quad : \forall{i}. \label{eq:refnode}
\end{align}
where $a_j$, $b_j$ and $c_j$ denote the generation cost coefficients and $p_{ji}$ represents the power output of the generator $j$ in hour $i$. $M_{k}^{I}$ represents the set of generators located at node $k$, $\hat{y}_{ki}$ indicates the forecasted load attached to node $k$ in hour $i$ and $f_{\ell{i}}$
corresponds to the DC power flow in the defined mapping set $A(\ell,k)$ with respect to the line $l$ and the node $k$. $\bar{p_{j}}$ is the maximum generation capacity for generator $j$. $R_{j}^{U}$ and $R_{j}^{D}$ denote the ramp-up and ramp-down limits of unit $j$, respectively, and $p_{j}^{ini}$ is the initial output of the generator $j$ at the beginning of the optimization problem. $B_{\ell}$ denotes the absolute value of susceptance of line $\ell$ and $\delta_{ki}$ stands for the voltage angle of node $k$ in hour $i$. 

Generally, the DAED model utilizes the hourly forecasted load $\hat{y}_{i}$ to generate the economic dispatch schedule for system operations for the next day. Here we use point forecasted load, a commonly applied approach by system operators in the DAED problem. If a stochastic approach is used, the model can be easily transformed to a stochastic model when probabilistic load is considered. Eq.~\eqref{eq:objective} indicates that the objective function for the DAED model is to minimize the total generation costs of all units. Using DC power flow, the power flow equations are incorporated via \eqref{eq:balance}, \eqref{eq:flowing} and \eqref{eq:transing}. Eqs.~\eqref{eq:active} to~\eqref{eq:rdinin} enforce the generation output and ramping limits. Lastly, \eqref{eq:ang} limits the range of the voltage angles, and \eqref{eq:refnode} defines the reference point for the power flow computation. 

\subsubsection{IPB model} 
\begin{align}
\begin{split}
\min_{\Tilde{p}_{ji},U_{\xi{i}}^{+},U_{\xi{i}}^{-},f_{\ell{i}},\delta_{ki}} \quad  \sum_{i=1}^{N}\sum_{j=1}^{J}\sum_{\xi=1}^{\Xi}{}&a_{j}\Tilde{p}_{ji}^{2}+b_{j}\Tilde{p}_{ji}+c_{j}+\\
&C_{\xi}^{+}\cdot{U_{\xi{i}}^{+}}+C_{\xi}^{-}\cdot{U_{\xi{i}}^{-}}
\end{split}\label{eq:obj_IFR}
\end{align}

\begin{align}
\sum_{j=1}^{J}{M_{k}^{I}\Tilde{p}_{ji}} &+ \sum_{\xi}^{\Xi}{M_{k}^{\Xi}}(U_{\xi{i}}^{+}-U_{\xi{i}}^{-})= y_{ki}+  \sum_{\ell=1}^{V}{A(\ell,k)f_{\ell{i}}}: \forall{k}, \forall{i}, \label{eq:balance_ac}    \\
& {p}_{ji} - RD_{ji} \leq \Tilde{p}_{ji} \leq {p}_{ji} + RU_{ji} \quad : \forall{j}, \forall{i}, \label{eq:reser}    \\
& 0 \leq U_{\xi{i}}^{+} \leq \bar{U}_{\xi}^{+}v_{\xi{i}}^{+} \quad : \forall{\xi}, \forall{i}, \label{eq:fa}\\
& 0 \leq U_{\xi{i}}^{-} \leq \bar{U}_{\xi}^{-}v_{\xi{i}}^{-} \quad : \forall{\xi}, \forall{i}, \label{eq:she}\\
& v_{\xi{i}}^{+} + v_{\xi{i}}^{-} = 1 \quad : \forall{\xi}, \forall{i}, \label{eq:bineq}\\
& v_{\xi{i}}^{+},v_{\xi{i}}^{-} \in \{0,1\} \quad : \forall{\xi}, \forall{i}, \label{eq:bindef}\\
& e_{\xi{i}} = e_{\xi(i-1)} + U_{\xi{i}}^{+} - U_{\xi{i}}^{-} \quad : \forall{\xi}, \forall{i}, \label{eq:ste}\\
& e_{\xi{i}} = E_{\xi}^{ini} \quad : i = 0, \forall{\xi}, \label{eq:e_ini}\\
& 0 \leq e_{\xi{i}} \leq e^{max}\quad : \forall{\xi}, \forall{i}, \label{eq:en_ran}\\
& \text{s.t. \eqref{eq:active} - \eqref{eq:refnode}.} \notag 
\end{align}
where $C_{\xi}^{+}$ and $C_{\xi}^{-}$, $U_{\xi{i}}^{+}$ and $U_{\xi{i}}^{-}$ denote the prices for positive and negative balancing services of online units and the charge and discharge power for the BESS unit $\xi$ in hour $i$, respectively. $M_{k}^{\Xi}$ indicates the set of BESS located at node $k$ and $y_{ki}$ is the actual load at node $k$ in hour $i$. $RD_{ji}$, $RU_{ji}$ denote the down/up reserve capacity for unit $j$ in hour $i$. $\Tilde{p}_{ji}$ is the adjusted generation output in the IPB process for unit $j$ in hour $i$. The BESS discharge and charge processes are bounded by $\bar{U}_{\xi}^{+}$ and $\bar{U}_{\xi}^{-}$. The binary variables $v_{\xi{i}}^{+}$ and $v_{\xi{i}}^{-}$ in constraint~\eqref{eq:bineq} ensure that the BESS $\xi$ cannot charge and discharge simultaneously. The variable $e_{\xi{i}}$ denotes the energy stored in BESS $\xi$ and $E_{\xi}^{ini}$ is the initial energy stored in the unit. 

When it comes to the intra-day process, the real-time load is likely to deviate from the day-ahead forecasts. Thus, the IPB process is necessary to maintain the system balance. The maximum balancing capacity, i.e., the sum of up/down reserve capacity and BESS discharge/charge capacity, is set to be larger than the largest hourly load deviation. As can be seen in the objective function~\eqref{eq:obj_IFR}, apart from the real-time generation cost, the costs of using the BESS services have to be included in the objective function. Meanwhile, the optimal power output of online units shall be adjusted according to the actual load, as shown in \eqref{eq:reser}, while satisfying the power flow equations \eqref{eq:balance_ac}. Additionally, the energy storage constraints with charge/discharge constraints for each BESS from \eqref{eq:fa} to~\eqref{eq:en_ran} shall be included in the optimization problem. It should be noted that the constraints from \eqref{eq:active} to~\eqref{eq:refnode} are still required to be included in the IPB optimization problem. With the formulated IPB problem, the real-time operation cost in hour $i$ can be determined as $C(y_i)$. 

Note that if the forecasting load is exactly the actual load in real-time, there is no need to adjust the power output nor utilizing BESS in the IPB process. Thus, using the DAED model alone, we can achieve the ideal cost $C(y_i^*)$ for a particular system. By comparing the costs between the ideal cost and the costs associated with multiple scenarios gained from the IPB model, the economic costs related to forecasting errors can thus be quantified.

\ifCLASSOPTIONcaptionsoff
\newpage
\fi
\end{document}